\documentclass[fleqn,usenatbib]{mnras}

\usepackage{newtxtext,newtxmath}
\usepackage{mathrsfs}
\usepackage[T1]{fontenc}
\usepackage{ae,aecompl}
\usepackage{booktabs}

\usepackage{graphicx}	% Including figure files
\usepackage{amsmath}	% Advanced maths commands
\usepackage{amssymb}	% Extra maths symbols

\title[$L_{\nu}$-$w_{\rm int, rest}$ relations and LFs of repeating and non-repeating FRBs]{Luminosity-duration relations and luminosity functions of repeating and non-repeating fast radio bursts}

\author[T. Hashimoto et al.]{
Tetsuya Hashimoto,$^{1,2}$\thanks{E-mail: tetsuya@phys.nthu.edu.tw}
Tomotsugu Goto,$^{1}$
Ting-Wen Wang,$^{1}$
Seong Jin Kim,$^{1}$
\newauthor
Simon C.-C. Ho,$^{1}$
Alvina Y. L. On,$^{1,2,3}$
Ting-Yi Lu, $^{1}$
and Daryl Joe D. Santos$^{1}$
\\
$^{1}$Institute of Astronomy, National Tsing Hua University, 101, Section 2. Kuang-Fu Road, Hsinchu, 30013, Taiwan (R.O.C.)\\
$^{2}$Centre for Informatics and Computation in Astronomy (CICA), National Tsing Hua University,\\101, Section 2. Kuang-Fu Road, Hsinchu, 30013, Taiwan (R.O.C.)\\
$^{3}$Mullard Space Science Laboratory, University College London, Holmbury St Mary, Surrey RH5 6NT, UK\\
}

\date{Accepted 2020 March 27. Received 2020 March 27; in original form 2019 November 16}

\pubyear{2020}

\begin{document}
\label{firstpage}
\pagerange{\pageref{firstpage}--\pageref{lastpage}}
\maketitle

\begin{abstract}
%just 250 now
Fast radio bursts (FRBs) are mysterious radio bursts with a time scale of approximately milliseconds.
Two populations of FRB, namely repeating and non-repeating FRBs, are observationally identified.
However, the differences between these two and their origins are still cloaked in mystery.
Here we show the time-integrated luminosity-duration ($L_{\nu}$-$w_{\rm int,rest}$) relations and luminosity functions (LFs) of repeating and non-repeating FRBs in the FRB Catalogue project.
These two populations are obviously separated in the $L_{\nu}$-$w_{\rm int,rest}$ plane with distinct LFs, i.e., repeating FRBs have relatively fainter $L_{\nu}$ and longer $w_{\rm int,rest}$ with a much lower LF.
In contrast with non-repeating FRBs, repeating FRBs do not show any clear correlation between $L_{\nu}$ and $w_{\rm int,rest}$.
These results suggest essentially different physical origins of the two.
%The faint ends of the luminosity functions of repeating and non-repeating FRBs are consistent with volumetric occurrence rates of soft gamma-ray repeaters (SGRs), type Ia supernovae, magnetars, and white-dwarf mergers, beyond neutron-star mergers and accretion-induced collapse (AIC) of white dwarfs.
The faint ends of the LFs of repeating and non-repeating FRBs are higher than volumetric occurrence rates of neutron-star mergers and accretion-induced collapse (AIC) of white dwarfs, and are consistent with those of soft gamma-ray repeaters (SGRs), type Ia supernovae, magnetars, and white-dwarf mergers.
This indicates two possibilities: either (i) faint non-repeating FRBs originate in neutron-star mergers or AIC and are actually repeating during the lifetime of the progenitor, or (ii) faint non-repeating FRBs originate in 
any of SGRs, type Ia supernovae, magnetars, and white-dwarf mergers.
The bright ends of LFs of repeating and non-repeating FRBs are lower than any candidates of progenitors, suggesting that bright FRBs are produced from a very small fraction of the progenitors regardless of the repetition.
Otherwise, they might originate in unknown progenitors.
\end{abstract}

\begin{keywords}
radio continuum: transients -- stars: magnetars -- stars: magnetic field -- stars: neutron -- (stars:) binaries: general -- stars: luminosity function, mass function
\end{keywords}

%%%%%%%%%%%%%%%%%%%%%%%%%%%%%%%%%%%%%%%%%%%%%%%%%%

%%%%%%%%%%%%%%%%% BODY OF PAPER %%%%%%%%%%%%%%%%%%

\section{Introduction}
\label{introduction}
Since the first discovery of a fast radio burst \citep[FRB; ][]{Lorimer2007}, $\sim$100 FRBs have been detected to date \citep[e.g., ][]{Petroff2016}.
There are two different types of FRBs: repeating and non-repeating FRBs.
The first repeating burst was named FRB 121102 \citep{Spitler2016}.
%The repeating radio bursts were first found for FRB 121102 \citep{Spitler2016}.
%Currently more than 100 repeats are detected for this FRB \citep{Spitler2016,Scholz2016,Spitler2018,Michilli2018,Zhang2018}.
The repeating signals of FRB 121102 have been confirmed more than 100 times until now \citep{Spitler2016,Scholz2016,Spitler2018,Michilli2018,Zhang2018}.
Recently, repeating FRBs have been discovered increasingly by the Canadian Hydrogen Intensity Mapping Experiment \citep[CHIME; ][]{CHIME1repeat2019,CHIMEFRB2019,CHIME8repeat2019} and the Robert C. Byrd Green Bank Telescope \citep[GBT; ][]{Kumar2019}.
In spite of the observational progress, the differences between repeating and non-repeating FRBs are still unclear because of observational limitations.
If a typical repeating time scale is longer than observational time scales or %repeats 
the luminosities of repeated bursts are too faint, such repeats can not be detected by current radio telescopes.
Thus, the FRBs may be mistakenly recognised as non-repeating FRBs.
A significant fraction of repeating FRBs might contaminate non-repeating FRBs.
In this sense, these two categories do not necessarily indicate two different origins.
%In fact, a volumetric occurrence rate of local non-repeating FRBs exceeds those of possible progenitor candidates, suggesting a repeating population in non-repeating FRBs \citep{Ravi2019repeat}.
In fact a volumetric occurrence rate, i.e., how many FRBs happen per unit time per unit volume, of %local 
nearby non-repeating FRBs exceeds those of possible progenitor candidates \citep{Ravi2019repeat}.
This suggests that at least some fractions of non-repeating FRBs originate from progenitors that emit multiple bursts over their lifetimes.

On the other hand, another FRB named FRB 171019 was originally considered as a non-repeating FRB.
However, two repetitions of radio bursts happened from this source and were reported recently \citep{Kumar2019}.
Therefore, a comparison of physical properties of repeating and non-repeating FRBs is one of the most important tasks to understand what makes this phenomenal difference.
Recently, \citet{CHIME8repeat2019} reported that the observed duration of repeating FRBs detected by CHIME is longer than that of non-repeating FRBs, suggesting different populations between repeating and non-repeating FRBs.

Theoretical models of the origins of FRBs are in chaos. 
So far $\sim$50 physical models of FRBs have been proposed \citep[e.g., ][]{Platts2019}.
There is no consensus yet on the physical origins of repeating and non-repeating FRBs.
There were some observational efforts to find FRBs at the positions of possible progenitors, e.g., remnants of super-luminous supernovae \citep{Law2019} and gamma-ray bursts \citep[GRBs; ][]{Madison2019,Men2019}.
However, there is so far no direct detection of FRBs from these possible progenitors.
Multi-wavelength and multi-messenger observations at the locations of FRBs were also conducted \citep[e.g., ][]{Callister2016,MAGIC2018,Sun2019,Martone2019,Tingay2019,Aartsen2020} with no clear detection of counterparts.

One way to constrain FRB origins is a luminosity-duration relation of FRBs. 
\citet{Hashimoto2019} found an unexpected positive correlation between the time-integrated luminosity, $L_{\nu}$, and rest-frame intrinsic duration, $w_{\rm rest, int}$, for non-repeating FRBs (see Sections \ref{calcLnu} and \ref{calcw_int_rest} for exact definitions of $L_{\nu}$ and $w_{\rm rest, int}$, respectively).
They argued that physical models which explicitly predict the correlation would be favoured for non-repeating FRBs.

In this paper, we compare repeating and non-repeating FRBs in the $L_{\nu}$-$w_{\rm int, rest}$ parameter space and in luminosity functions.
The structure of the paper is as follows:
we describe a compilation of our FRB sample in Section \ref{sample}.
In Section \ref{analysis}, we demonstrate calculations of $L_{\nu}$, $w_{\rm int, rest}$, and luminosity functions.
Results of the $L_{\nu}$-$w_{\rm int, rest}$ relations and luminosity functions are described in Section \ref{results}.
The implications of our results on repeating and non-repeating FRB populations and their origins are discussed in Section \ref{discussion} followed by conclusions in Section \ref{conclusion}.
Throughout the paper, we assume the {\it Planck15} cosmology \citep{Planck15} as a fiducial model, i.e., $\Lambda$ cold dark matter cosmology with ($\Omega_{m}$,$\Omega_{\Lambda}$,$\Omega_{b}$,$h$)=(0.307, 0.693, 0.0486, 0.677), unless otherwise mentioned.

\section{Sample}
\label{sample}
We compiled 90 \lq verified\rq\ FRBs from the FRB Catalogue (FRBCAT) project\footnote[1]{\url{http://frbcat.org/}} \citep{Petroff2016} as of 21 August 2019, which were confirmed through publication, or received with high importance scores in the VOEvent Network.
The original FRBCAT catalogue contains FRB ID, telescope, galactic latitude ($b$), longitude ($l$), sampling time ($w_{\rm sample}$), central frequency ($\nu_{\rm obs}$), observed dispersion measure (DM$_{\rm obs}$), observed burst duration ($w_{\rm obs}$), and observed fluence ($E_{\nu_{\rm obs}}$) together with errors of these observed parameters.
In cases where the $E_{\nu_{\rm obs}}$ error, $\delta E_{\nu_{\rm obs}}$, is not provided in literature, we calculated it using
%as follows.
\begin{equation}
\delta E_{\nu_{\rm obs}}= E_{\nu_{\rm obs}}\{(\delta w_{\rm obs}/w_{\rm obs})^{2}+(\delta F_{\rm obs}/ F_{\rm obs})^{2} \}^{1/2},
\end{equation}
where $\delta w_{\rm obs}$, $\delta F_{\rm obs}$, and $F_{\rm obs}$ are $w_{\rm obs}$ error, observed flux density error, and observed flux density, respectively.
%If either of $\delta w_{\rm obs}$ or $\delta F_{\rm obs}$ is not provided in literature, we assumed 10\% uncertainty, i.e., $\delta w_{\rm obs}/w_{\rm obs}=0.1$ or $\delta F_{\rm obs}/F_{\rm obs}=0.1$.
%We adopted these values since median values of $\delta w_{\rm obs}/w_{\rm obs}$ and signal-to-noise ratios of the detected bursts are $\sim$0.1 and $\sim$16, respectively, in our sample.

If $\delta w_{\rm obs}$ is not provided in literature, we assumed 10\% uncertainty, i.e., $\delta w_{\rm obs}/w_{\rm obs}=0.1$.
We adopted this value because the median of $\delta w_{\rm obs}/w_{\rm obs}$ is $\sim$0.1 in our sample.

If $\delta F_{\rm obs}$ is not given in literature, we assumed $\delta F_{\rm obs}$ to be telescope-dependent. 
Among 90 verified FRBs catalogued in FRBCAT, 10 FRBs do not have $\delta F_{\rm obs}$: 
%Such FRBs include 
one Arecibo, three Pushchino, four UTMOST, one CHIME, and one DSA-10 FRBs. 
The assumed fractional uncertainties, $\delta F_{\rm obs}/F_{\rm obs}$, are 6, 0.3, and 0.5 for Arecibo, UTMOST, and CHIME, respectively.
These values are empirically determined for individual telescopes by calculating median values of $\delta F_{\rm obs}/F_{\rm obs}$ reported in the FRBCAT project. 
We caution readers that these fractional uncertainties are subjected to various uncertainties. 
For example, if the slope of the source counts is much flatter than $\alpha_{\rm SC}\lesssim1.1$, the deviations from the published values might be large. 
However, since the accurate value of $\alpha_{\rm SC}$ is still unknown, the empirically estimated uncertainties might be updated once more accurate measurements become available. 
The fluence uncertainties of FRBs detected with Pushchino are not explicitly reported \citep{Fedorova2019}.
We assumed a conservative fractional uncertainty of 0.5 for Pushchino FRBs, while the signal-to-noise ratios of the FRBs are 6.2, 9.1, and 8.3 \citep{Fedorova2019}. 
These signal-to-noise ratios of Pushchino FRBs are systematically lower than those of FRBs detected with other telescopes.
In this regard, Pushchino FRBs are not used in calculating the luminosity functions in Section \ref{calc_LF}.
The fluence uncertainty of DSA-10 is assumed to be 10\% since the observed fluence exceeded 8$\sigma$ detection limit by $\sim15$\% with an accurate localisation within the field of view \citep{Ravi2019}.

The reported fluences of FRBs detected with Parkes, CHIME and UTMOST are actually lower limits because the burst locations with respect to the beam centre are unknown.
Individual fluence corrections for the positional uncertainties are difficult for such FRBs. 
This uncertainty is statistically included in the calculations of the luminosity function of each telescope in Section \ref{calc_LF}. 
We do not include this uncertainty in the ASKAP luminosity function because ASKAP is not affected by this uncertainty.

Intra-channel bandwidth, $\Delta \nu_{\rm obs}$, is compiled from references therein, which is necessary for calculation of dispersion smearing.
Spectral index, $\alpha$, is also compiled from references therein if available, otherwise we assumed a mean value of $\alpha=-1.5$ \citep{Macquart2019}, where $E_{\nu_{\rm obs}} \propto \nu_{\rm obs}^{\alpha}$.

In order to treat irregular cases, we added four flags in the catalogue including \lq scattering flag\rq, \lq repeating flag\rq, \lq intrinsic-duration flag\rq, and \lq spec-$z$ flag\rq.
The scattering flag indicates the contamination of scattering broadening to $w_{\rm obs}$.
We rely on the estimates in literature and FRBCAT to determine the scattering flag.
This flag is off when $w_{\rm obs}$ is reported after the deconvolution of the scattering tail \citep[e.g.,][]{Shannon2018,CHIMEFRB2019,CHIME8repeat2019}.
The repeating flag indicates confirmed repeating FRBs \citep[e.g., ][]{Spitler2016,CHIME1repeat2019,CHIME8repeat2019,Kumar2019}.
The intrinsic-duration flag is on if the reported burst duration is already corrected for instrumental and scattering broadening effects \citep{CHIMEFRB2019}.
In such cases, we adopt the reported duration as an intrinsic duration, $w_{\rm int}$, instead of our calculation.
The spec-$z$ flag is for four FRBs with spectroscopic redshift measured from the host galaxies \citep[FRB 121102, 180916.J0158+65, 180924, and 190523; ][]{Tendulkar2017,Marcote2020,Bannister2019,Ravi2019}.
For these four FRBs we use the spectroscopic redshifts instead of the redshift estimated from the dispersion measure.
Recently another host galaxy was identified \citep[FRB 181112 host; ][]{Prochaska2019}.
This burst is not included in this work, since the burst information has not been listed in the FRBCAT as of 21 August 2019.

In the current FRBCAT, some of the individual bursts of repeating FRBs are incomplete. 
To supply the missing bursts, we compiled all of the parameters described above for each burst of repeating FRBs reported in other literature, i.e., 11 Arecibo repeats \citep{Spitler2016}, 5 GBT and 1 Arecibo \citep{Scholz2016}, and 93 GBT \citep{Zhang2018} for FRB 121102, 6 CHIME repeats for FRB 180814.J0422+73 \citep{CHIMEFRB2019,CHIME1repeat2019}, 2-10 CHIME repeats for 8 FRBs \citep{CHIME8repeat2019}, and 1 ASKAP and 2 GBT repeats for FRB 171019 \citep{Kumar2019}.

If spec-$z$ is not available, we excluded FRBs with dispersion measures dominated by the Milky Way and halo contribution, i.e., DM$_{\rm obs}-$DM$_{\rm MW}-$DM$_{\rm halo} \leqq 0$, where DM$_{\rm MW}$ is a dispersion measure contributed by the interstellar medium in the Milky Way and DM$_{\rm halo}$ is a contribution from the dark matter halo hosting the Milky Way (see Section \ref{analysis} for details).
This is because the uncertainties of redshift and distance are too large to calculate the luminosity. 

After applying this criterion, our sample includes a total of 11 repeating FRBs with 144 repeats and 77 non-repeating FRBs. 

\section{Analysis}
\label{analysis}
\subsection{Time-integrated luminosity}
\label{calcLnu}
%We followed \citet{Hashimoto2019} to calculate a time-integrated luminosity and rest-frame intrinsic duration of FRBs based on our sample.
We calculated a time-integrated luminosity and rest-frame intrinsic duration of our FRB sample in a similar way to \citet{Hashimoto2019}.
Here we briefly describe the process.
We assumed YMW16 electron-density model \citep{Yao2017} to estimate DM$_{\rm MW}$. 
The DM$_{\rm MW}$ was accumulated up to 10 kpc along the line of sight to FRBs.
Dispersion measures contributed from FRB host galaxies, DM$_{\rm host}$, were parameterised as DM$_{\rm host}=50.0/(1+z)$ pc cm$^{-3}$ 
%by
following 
%literature 
\citet{Shannon2018}.
There are several studies in which DM$_{\rm halo}$ are investigated \citep[e.g., ][]{Dolag2015,Prochaska2019DMhalo,Keating2020}.
We assumed DM$_{\rm halo}=65$ pc cm$^{-3}$, which is a mean between 50 and 80 pc cm$^{-3}$
%reported by 
\citep{Prochaska2019DMhalo}.
After subtracting DM$_{\rm MW}$, DM$_{\rm halo}$, and DM$_{\rm host}$ from DM$_{\rm obs}$, the remaining term is a contribution from inter-galactic medium, DM$_{\rm IGM}$.
While DM$_{\rm IGM}$ should fluctuate along different line of sights, the mean value of DM$_{\rm IGM}$ is expressed as a function of redshift with some cosmological parameters \citep[e.g., ][]{Zhou2014}.
By assuming the cosmological parameters, redshift is roughly estimated from DM$_{\rm IGM}(z)$ with uncertainties originated from the line-of-sight fluctuation of DM$_{\rm IGM}$ and DM$_{\rm obs}$ error, $\delta$DM$_{\rm obs}$.
As a conservative estimate, we adopted the highest uncertainty of DM$_{\rm IGM}(z)$, $\delta$DM$_{\rm IGM} (z)$, among simulations with different resolutions \citep{Zhu2018}.
We used spectroscopic redshift instead of DM-derived redshift if available \citep[i.e., FRB121102, 180916.J0158+65, 180924, and 190523: ][]{Tendulkar2017,Marcote2020,Bannister2019,Ravi2019}.
We calculated time-integrated luminosities of FRBs at rest-frame 1.83 GHz by using Eq. 6 in \citet{Hashimoto2019}.
This rest-frame frequency is selected so that the $K$-correction term in Eq. 6, $\frac{1}{(1+z)^{2+\alpha}} \left( \frac{\nu_{\rm rest}}{\nu_{\rm obs}}\right)^{\alpha}$, can be minimised.
We note that a spectral index, $\alpha=-1.5$ \citep{Macquart2019}, is assumed for FRBs without $\alpha$ measurement.

\subsection{Rest-frame intrinsic duration}
\label{calcw_int_rest}
In general, FRB pulse duration is broadened by dispersion smearing, data sampling time interval, and scattering.
The dispersion smearing is caused by a finite spectral resolution of the instrument.
The pulse delay within the observational spectral resolution broadens the observed duration.
The instrumental sampling time also causes pulse broadening.
To derive the intrinsic duration of FRBs ($w_{\rm int}$), we removed these two instrumental effects on pulse broadening by following Eq. 5 in \citet{Hashimoto2019}.
In the case when $w_{\rm int}$ is not instrumentally resolved, i.e., $w_{\rm obs}^{2} \leqq w_{\rm sample}^{2}+w_{\rm DM}^{2}$, we use $w_{\rm obs}$ as the upper limit of $w_{\rm int}$, where $w_{\rm DM}$ is the pulse broadening by dispersion smearing.
Note that scattering broadening is not explicitly removed in this work except for some cases in which deconvoluted duration is reported in the literature \citep[e.g., ][]{Shannon2018,CHIMEFRB2019,CHIME8repeat2019}.
Instead of the deconvolution, we flagged the scattering feature.
Therefore, the intrinsic duration of FRBs with scattering flags could be shorter than the reported values.
The rest-frame intrinsic duration, $w_{\rm int, rest}$, was calculated as $w_{\rm int, rest}=w_{\rm int}/(1+z)$.

\begin{figure*}
	\includegraphics[width=6.35in]{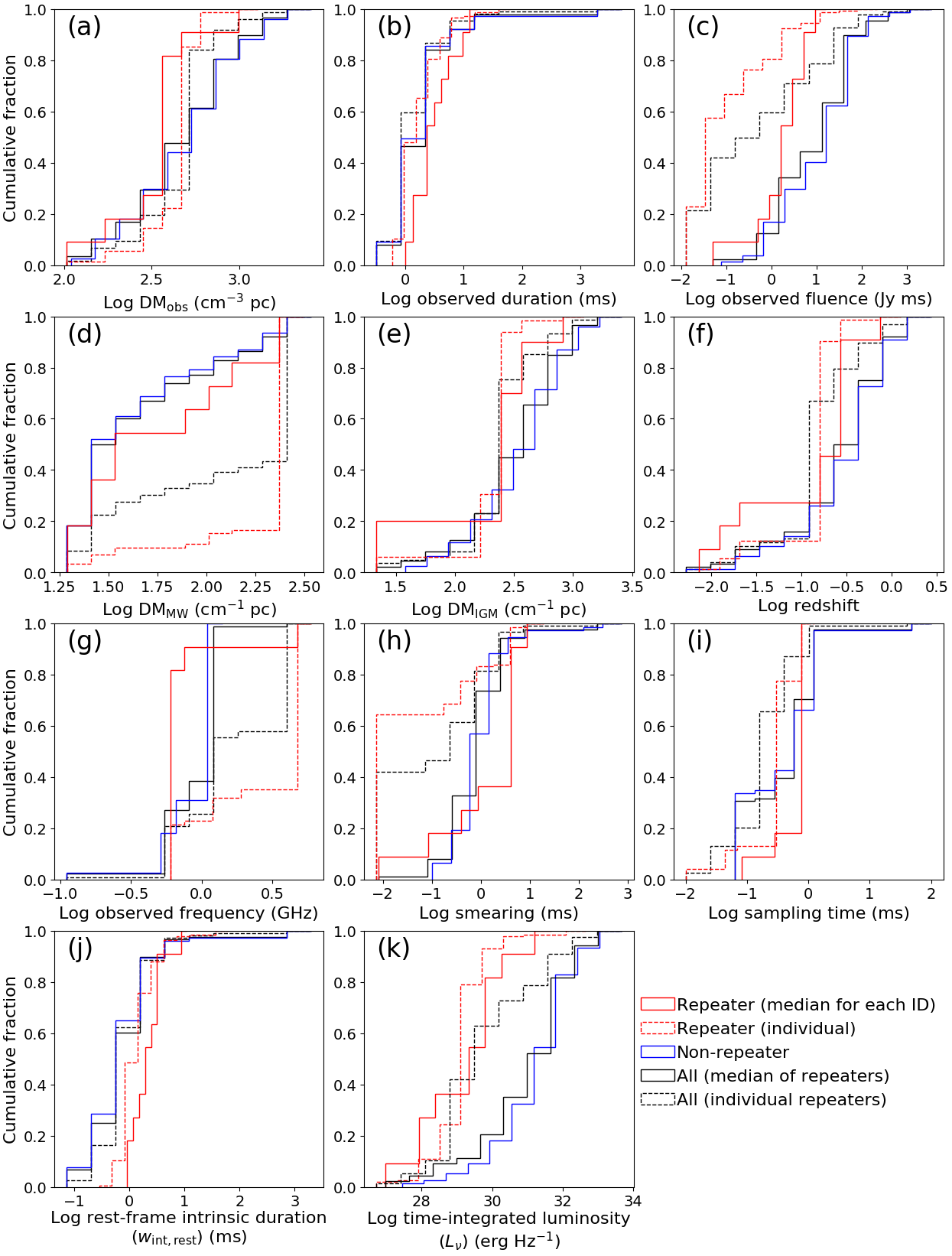}
    \caption{Cumulative fractions as a function of parameters of our FRB sample.
    From (a) to (k), the cumulative histograms are for observed dispersion measure, observed duration, observed fluence, dispersion measure of the Milky Way, dispersion measure of inter-galactic medium, redshift, observed frequency, dispersion smearing, instrumental sampling time, rest-frame intrinsic duration, and time-integrated luminosity at rest-frame 1.83 GHz.
    Black, red, and blue lines correspond to the total sample, repeating FRBs, and non-repeating FRBs, respectively.
    Solid black and red lines include repeating FRBs counted such that an identical FRB ID is single source with median values of parameters among the repeats, e.g., FRB 121102 is counted as one FRB.
    Dashed lines include repeats individually counted for each repeating FRB, e.g., FRB 121102 is counted as 110 FRBs.
    Note that upper limits are included in the histogram of the (j) rest-frame intrinsic duration.
    Cumulative histograms of DM$_{\rm IGM}$ and redshift for each telescope are shown in Figs. \ref{figA1} and \ref{figA2}, respectively, in Appendix A.
    }
    \label{fig1}
\end{figure*}

Histograms of cumulative fractions of observed, instrumental, and derived parameters of our sample are summarised in Fig. \ref{fig1}.
In the histograms, each sample is binned into 10 subsamples ranging from the minimum and maximum values.
Cumulative histograms of DM$_{\rm IGM}$ and redshift for each telescope are shown in Figs. \ref{figA1} and \ref{figA2}, respectively, in Appendix A.
Monte Carlo simulations were performed to calculate errors of $L_{\nu}$ and $w_{\rm int,rest}$ by independently assigning random 10,000 errors to DM$_{\rm obs}$, $w_{\rm obs}$, DM$_{\rm IGM}(z)$, and $E_{\nu_{\rm obs}}$.
Here, the line-of-sight fluctuation of DM$_{\rm IGM}$ mentioned above is included in the error of DM$_{\rm IGM}(z)$.
Each random error is assumed to follow a Gaussian probability distribution function with a standard deviation of the observational uncertainty.

\subsection{Rest-frame cadences of repeating FRBs}
We also calculated rest-frame cadences of repeating FRBs.
The rest-frame cadence is calculated as $N_{\rm repeat}(1+z)/t_{\rm obs}$, where $N_{\rm repeat}$ is the number of repeats of each FRB and $t_{\rm obs}$ is the observed time on source. The values of $N_{\rm repeat}$, $t_{\rm obs}$, and references of 11 repeating FRBs in our sample are summarised in Table \ref{tab1}.

\begin{table*}
	\centering
	\caption{
	Number of repeats of each repeating FRB and observed time.
	}
	\label{tab1}
	\begin{flushleft}
	\begin{tabular}{|l|c|c|c|}\hline
 ID & $N_{\rm repeat}$ & $t_{\rm obs}$ & Reference \\ 
    &                  &     (hour)      & \\ \hline
121102 & 110  &  84.1 &  \citet{Scholz2016,Zhang2018}  \\ 
171019 & 3    &  1009.6 & \citet{Kumar2019} \\
180814.J0422+73 & 6 & 23 & \citet{CHIME1repeat2019,CHIMEFRB2019} \\
180916.J0158+65 & 10 & $23\pm 8$ & \citet{CHIME8repeat2019} \\
181017.J1705+68 & 2 & $20 \pm 11$ & \citet{CHIME8repeat2019} \\
181030.J1054+73 & 2 & $46 \pm 18$ & \citet{CHIME8repeat2019} \\
181119.J12+65 & 3 & $19 \pm 9$ & \citet{CHIME8repeat2019} \\
181128.J0456+63 & 2 & $16 \pm 10$ & \citet{CHIME8repeat2019} \\
190116.J1249+27 & 2 & $8 \pm 5$ & \citet{CHIME8repeat2019} \\
190209.J0937+77 & 2 & $62 \pm 26$ & \citet{CHIME8repeat2019} \\
190222.J2052+69 & 2 & $20 \pm 10$ & \citet{CHIME8repeat2019} \\ \hline
    \end{tabular}\\
    \end{flushleft}
\end{table*}

\subsection{Calculation of luminosity function}
\label{calc_LF}
\begin{figure*}
	\includegraphics[width=\columnwidth]{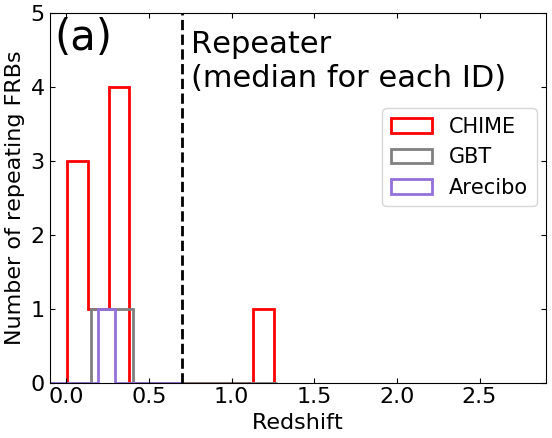}
	\includegraphics[width=\columnwidth]{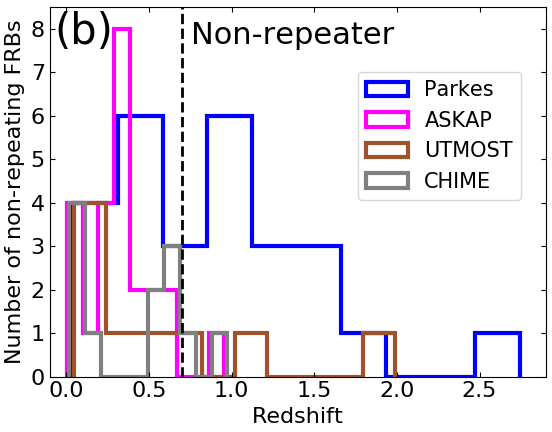}
    \caption{
    (Left) Redshift distributions of repeating FRBs detected by CHIME, GBT, and Arecibo.
    The number of repeating FRBs are counted such that the identical FRB ID is the single source.
    Here we use median redshift of repeats for each repeating FRBs.
    Dashed vertical line, $z=0.7$, is the redshift cutoff applied to calculations of the luminosity functions.
    (Right) Same as left except for non-repeating FRBs detected by Parkes, ASKAP, UTMOST, and CHIME.
    All of our sample is shown before applying the redshift cuts and detection limits for the calculations of luminosity functions.
    }
    \label{fig2}
\end{figure*}

Here, we calculate the FRB luminosity functions for each telescope, because different telescopes have different sensitivities and survey volumes. 
For this purpose we divided our sample into subsamples detected with the same telescope.
We only consider FRBs at $0.01\leqq z<0.7$ to calculate the luminosity functions.
The upper limit could mitigate a possible volumetric density evolution of progenitors with redshift (e.g., Hashimoto et al. 2020b in prep.).
The lower limit is applied to exclude very close events that could involve huge uncertainties on the DM-derived distances (see also discussion in Section \ref{FRBmodel}).
The upper limit is shown by a dashed vertical line in Fig. \ref{fig2}.
Table \ref{tab2} and \ref{tab3} summarise the number of FRBs (repeating and non-repeating, respectively) observed with each telescope in our sample.
The CHIME detections of repeating FRBs are used to derive the luminosity function of repeating FRBs.
The repeating FRBs observed with GBT and Arecibo are also included in our analysis to constrain the upper limit of the luminosity function.
As for non-repeating FRBs, only Parkes, ASKAP, CHIME, and UTMOST observations are considered to derive luminosity functions because of the small statistics of other telescopes.

To estimate the luminosity functions of FRBs, we use a simple $V_{\rm max}$ method \citep[e.g., ][]{Schmidt1968,Avni1980}, where the 4$\pi$ coverage of $V_{\rm max}$ ($V_{\rm max,4\pi}$) is expressed as
\begin{equation}
\label{eq2}
V_{\rm max,4\pi}=\frac{4\pi}{3}(d_{\rm max}^{3}-d_{\rm min}^{3}).
\end{equation}
Here $d_{\rm min}$ is a comoving distance to $z=0.01$, which is the lower redshift limit applied to our sample.
$d_{\rm max}$ is a maximum comoving distance for a FRB with a time-integrated luminosity, $L_{\nu_{\rm rest}}$, to be detected with a specific fluence limit, $E_{\rm lim}$.
In previous studies, $E_{\rm lim}$ is reported in terms of $E_{\nu}$ for Parkes and UTMOST \citep{Keane2015,Caleb2016} and in terms of $E_{\nu}\times w_{\rm obs}^{1/2}$ for ASKAP, CHIME, and Arecibo \citep{Shannon2018,CHIMEFRB2019,Spitler2014}.
To avoid systematic offsets in the luminosity functions due to the different definitions of $E_{\rm lim}$, we empirically derived $E_{\rm lim}$ for each telescope with the same definition described in Appendix B. %which is approximated by a minimum fluence of FRB detected with each telescope.
The value of $E_{\rm lim}$ is approximated by the peak of the data distributed along the perpendicular direction to the $w_{\rm obs}^{1/2}$ dependency in the $E_{\nu_{\rm obs}}$-$w_{\rm obs}$ space, such that the duration dependency can be taken into account.
The data distribution and histograms in the $E_{\nu_{\rm obs}}$-$w_{\rm obs}$ space of our sample are shown in Figs. \ref{figB1} and \ref{figB2}.
The adopted $E_{\rm lim}$ are summarised in Tables \ref{tab2} and \ref{tab3}.
By using Eq. 6 in \citet{Hashimoto2019}, $d_{\rm max}$ is expressed as
\begin{equation}
\label{eq3}
d_{\rm max} (z_{\rm max})=\left\{\frac{L_{\nu_{\rm rest}}(1+z_{\rm max})^{2+\alpha}}{4\pi E_{\rm lim}}\left(\frac{\nu_{\rm obs}}{\nu_{\rm rest}}\right)^{\alpha}\right\}^{1/2}(1+z_{\rm max})^{-1},
\end{equation}
where $z_{\rm max}$ is redshift at the comoving distance of $d_{\rm max}$.
We adopt the rest-frame frequency, $\nu_{\rm rest}$, at 1.83 GHz \citep{Hashimoto2019}.
Since the left term of Eq. \ref{eq3}, comoving distance, is calculated with a cosmological assumption, the solution to $z_{\rm max}$ of the Eq. \ref{eq3} provides individual FRBs with $d_{\rm max}$.
Note that we adopted $z_{\rm max}=0.7$ if the solution is higher than 0.7, 
%because we limit our sample to FRBs at $0.01\leqq z<0.7$.
so that $z_{\rm max}$ can not exceed the redshift cut, $z<0.7$.
Based on Eq. \ref{eq2} and \ref{eq3}, we calculated $V_{\rm max,4\pi}$ of individual FRBs.

Each FRB was detected in a comoving volume of $V_{\rm max,4\pi}\times \Omega_{\rm sky}$ during rest-frame survey time, $t_{\rm rest}=t_{\rm obs}/(1+z_{\rm FRB})$, where $\Omega_{\rm sky}$ and $z_{\rm FRB}$ are a fractional sky coverage of the survey and redshift of the FRB, respectively.
Therefore, the number density of each FRB per unit time, $\rho (L_{\nu_{\rm rest}})$, is
\begin{equation}
\label{eq4}
\rho(L_{\nu_{\rm rest}})=1/(V_{\rm max,4\pi}\Omega_{\rm sky}t_{\rm rest})=(1+z_{\rm FRB})/(V_{\rm max,4\pi}\Omega_{\rm sky}t_{\rm obs}).
\end{equation}
Each survey provides a different value of $\Omega_{\rm sky}t_{\rm obs}$.
When FRBs are detected via multiple surveys, $\Omega_{\rm sky}t_{\rm obs}$ was accumulated for each telescope as follows.
\begin{equation}
\label{eq5}
\rho(L_{\nu_{\rm rest}})=(1+z_{\rm FRB})/(V_{\rm max,4\pi}\Sigma_{i}\Omega_{\rm sky, {\it i}}t_{\rm obs, {\it i}}),
\end{equation}
where $i$ denotes $i$th survey with the same telescope.
The adopted $E_{\rm lim}$, $\Omega_{\rm sky, {\it i}}t_{\rm obs, {\it i}}$, and their references are summarised in Table \ref{tab2} and \ref{tab3}.

\begin{figure*}
	\includegraphics[width=2.25in]{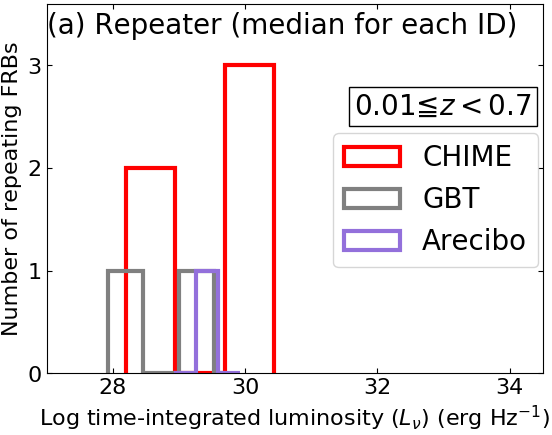}
	\includegraphics[width=2.25in]{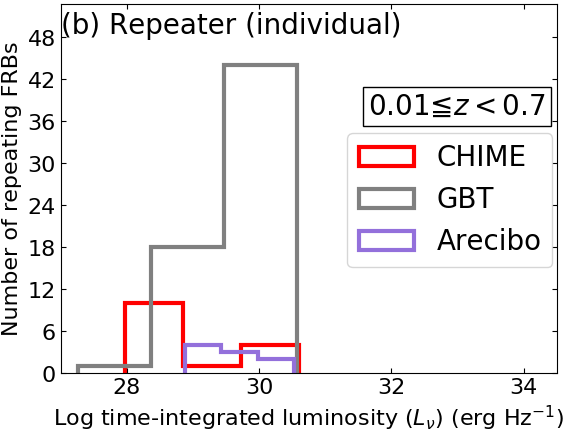}
	\includegraphics[width=2.25in]{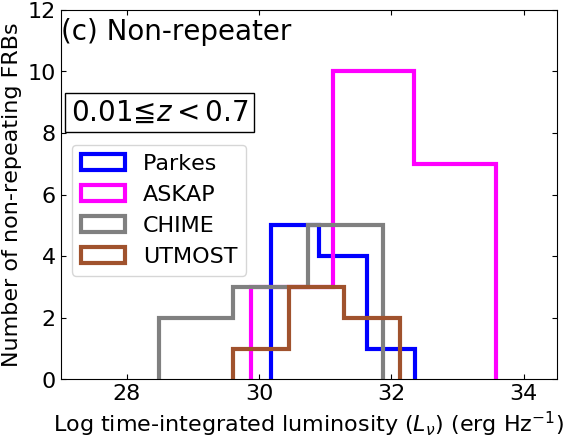}
    \caption{(Left) Histograms of the time-integrated luminosities at rest-frame 1.83 GHz of repeating FRBs.
    Repeating FRBs are counted such that the identical FRB ID is the single source.
    (Middle) Same as left except for counting repeats individually for each repeating FRB.
    (Right) Same as left except for non-repeating FRBs.
    Only FRBs used for the calculations of luminosity functions are demonstrated.
    }
    \label{fig3}
\end{figure*}

Each telescope sample was divided into three luminosity bins, $L_{j} (j=1,2,3)$, in logarithmic scale (Fig. \ref{fig3}). 
Here we use the three bins in order to secure a meaningful number of samples in each bin.
Within the luminosity bins, $\rho (L_{j})$ is summed to derive luminosity function, $\Phi$, i.e., 
\begin{equation}
\Phi (L_{j})=\Sigma_{k}\rho(L_{j,k})/\Delta\log L,
\end{equation}
where the subscript $k$ denotes the $k$th FRB in $L_{j}$ bin and $\Delta\log L$ is the luminosity bin size.
Note that the number of luminosity bins for CHIME/GBT and Arecibo repeating FRBs (median case) are adopted to be two and one, respectively, due to their small statistics.

With respect to the unknown positions of FRBs within the fields of view of Parkes, CHIME and UTMOST, \citet{Macquart2018} reported that the averaged fluence can vary by a factor of 1.7 depending on the slope of the source counts, $\alpha_{\rm SC}$.
Considering that the reported fluences for Parkes, CHIME, and UTMOST are lower limits, we included this factor as a systematic uncertainty of fluence and thus time-integrated luminosity in the luminosity functions.
The effective survey area in Table \ref{tab3} also depends on $\alpha_{\rm SC}$ \citep{Macquart2018}.
In order to accurately estimate the dependency, $\alpha_{\rm SC}$ and beam pattern have to be correctly calculated for each telescope. 
Such extensive analysis is out of scope of this paper.
\citet{Macquart2018} reported that the effective survey area can increase by a factor of $\sim3$ depending on $\alpha_{\rm SC}$.
Instead of carrying out an extensive analysis, we included the factor of 3 as a systematic uncertainty on the survey area when the luminosity functions are calculated. 

In summary, we used FRBs satisfying the following criteria for the calculations of luminosity functions:
\begin{itemize}
%\item DM$_{\rm obs}-$DM$_{\rm MW}-$DM$_{\rm halo} > 0$ unless spec-$z$ is available
\item $0.01\leqq$ redshift (spec-$z$ if available) $<0.7$
\item $E_{\nu_{\rm obs}} \geqq E_{\rm lim}$
\end{itemize}
The sample for luminosity functions includes a total of 7 repeating FRBs with 87 repeats and 46 non-repeating FRBs.

%\begin{figure}
%	% To include a figure from a file named example.*
%	% Allowable file formats are eps or ps if compiling using latex
%	% or pdf, png, jpg if compiling using pdflatex
%	\includegraphics[width=\columnwidth]{figure3.png}
%    \caption{
%    Fluence distributions of FRBs detected by Parkes, ASKAP, UTMOST, CHIME, GBT, and Arecibo.
%    Both repeating and non-repeating FRBs are included.
%    The number of repeats of repeating FRBs are individually counted for each repeating FRB.
%    }
%    \label{fig3}
%\end{figure}

\begin{table*}
	\centering
	\caption{
	Total number of repeating FRBs observed with each telescope in our sample and individual survey parameters.}
	\label{tab2}
	\begin{flushleft}
	\begin{tabular}{|l|c|c|c|c|c|}\hline
 &  & \multicolumn{3}{c}{Repeating FRBs} \\ \hline
Telescope  &Number$^{a}$&Number$^{a}$ &$E_{\rm lim}$ $^{d}$& $\Omega_{\rm sky, {\it i}}t_{\rm obs, {\it i}}$ $^{e}$&$\Omega_{\rm sky, {\it i}}t_{\rm obs, {\it i}}$ reference \\ 
    & (all) & ($0.01\leqq z<0.7$ and $E_{\nu_{\rm obs}} \geqq E_{\rm lim}$) &(Jy ms)         & (deg$^{2}$ hour) &                    \\ \hline
%CHIME  & 9 (31)& 5 (15)$^{c}$ & 1.67$w_{\rm obs}^{1/2}$  & $1.39\times10^{6}$& \citet{CHIME8repeat2019,CHIME1repeat2019} \\
CHIME  & 9 (31)& 5 (15)$^{c}$ & 1.7$w_{\rm obs}^{1/2}$  & $1.39\times10^{6}$& \citet{CHIME8repeat2019,CHIME1repeat2019} \\
GBT$^{b}$    &  2 (100)& 2 (63)$^{c}$ & 0.032$w_{\rm obs}^{1/2}$ & 1.25 & \citet{Scholz2016} \\        
       &     &        & & 0.28 & \citet{Zhang2018}  \\
       &     &        & & 0.58 & \citet{Kumar2019} \\ 
Arecibo$^{b}$ & 1 (12) & 1 (9)$^{c}$ & 0.052$w_{\rm obs}^{1/2}$ & 6.23 & \citet{Spitler2014} \\
       &       &      & & 0.23 & \citet{Spitler2016} \\
ASKAP$^{b}$  & 1 (1) & 1(1) &   &      & \\ \hline
    \end{tabular}\\
    $^{a}$ Repeating FRBs are counted such that the identical FRB ID is the single source. Numbers in parentheses are individual counts of repeats. Note that a few repeating FRBs were detected in multiple telescopes.
    $^{b}$ Observations were targeted to repeating FRBs, which places upper limits on FRB number densities and luminosity functions.
    $^{c}$ FRBs used for the calculations of luminosity functions. 
    $^{d}$ Detection limit, $E_{\rm lim}$, is approximated by a peak of data distribution along the perpendicular direction to the $w_{\rm obs}^{1/2}$ (ms$^{1/2}$) dependency in the $E_{\nu_{\rm obs}}$-$w_{\rm obs}$ space (see Appendix B).
    $^{e}$ Survey area times exposure time on sky.
    %a minimum fluence of FRB detected with each telescope.
    \end{flushleft}
\end{table*}

\begin{table*}
	\centering
	\caption{
	Total number of non-repeating FRBs observed with each telescope in our sample and individual survey parameters.}
	\label{tab3}
	\begin{flushleft}
	\begin{tabular}{|l|c|c|c|c|c|}\hline
 & & \multicolumn{3}{c}{Non-repeating FRBs}  \\ \hline
Telescope  &Number &Number &$E_{\rm lim}$ $^{b}$& $\Omega_{\rm sky, {\it i}}t_{\rm obs, {\it i}}$ &$\Omega_{\rm sky, {\it i}}t_{\rm obs, {\it i}}$ reference             \\ 
           &(all) &($0.01\leqq z<0.7$ and $E_{\nu_{\rm obs}}\geqq E_{\rm lim}$)&(Jy ms)         & (deg$^{2}$ hour) &                    \\ \hline
Parkes     &27 &10$^{a}$ &0.72$w_{\rm obs}^{1/2}$           & 267    &\citet{Zhang2019} \\ 
           &         &     &         & 4394.5 &\citet{Oslowski2019} \\
%ASKAP      &24 &20$^{a}$ &28.2$w_{\rm obs}^{1/2}$          & $1.32\times10^{4}$ &\citet{Bannister2017}\\
ASKAP      &24 &20$^{a}$ &28$w_{\rm obs}^{1/2}$          & $1.32\times10^{4}$ &\citet{Bannister2017}\\
           &        &      &         & 5.1$\times$10$^5$ & \citet{Shannon2018} \\
           &        &      &         & $1.36\times10^{4}$ &\citet{Macquart2019} \\   
           &        &      &         & 255    &\citet{Bannister2019} \\
%CHIME      &12&10$^{a}$ &1.67$w_{\rm obs}^{1/2}$           & $<2.83\times10^{4}$&\citet{CHIMEFRB2019} \\
CHIME      &12&10$^{a}$ &1.7$w_{\rm obs}^{1/2}$           & $<2.83\times10^{4}$&\citet{CHIMEFRB2019} \\
%UTMOST     &9 &6$^{a}$  &12.9$w_{\rm obs}^{1/2}$          & $2.58\times10^{4}$ &\citet{Caleb2017} \\
UTMOST     &9 &6$^{a}$  &13$w_{\rm obs}^{1/2}$          & $2.58\times10^{4}$ &\citet{Caleb2017} \\
           &        &      &         & $7.43\times10^{4}$ &\citet{Farah2019} \\
Pushchino  &2       &-      &        &                   &  \\
DSA-10     &1       &-      &        &                   & \\
GBT        &1       &1      &        &                   & \\
Arecibo    &1       &1      &        &                   & \\ \hline
    \end{tabular}\\
    $^{a}$ FRBs used for the calculations of luminosity functions. 
    $^{b}$ Detection limit, $E_{\rm lim}$, is approximated by a peak of data distribution along the perpendicular direction to the $w_{\rm obs}^{1/2}$ (ms$^{1/2}$) dependency in the $E_{\nu_{\rm obs}}$-$w_{\rm obs}$ space (see Appendix B).
    %a minimum fluence of FRB detected with each telescope.
    \end{flushleft}
\end{table*}

\section{Results}
\label{results}
\subsection{Luminosity-duration relation}
\begin{figure*}
	\includegraphics[width=\columnwidth]{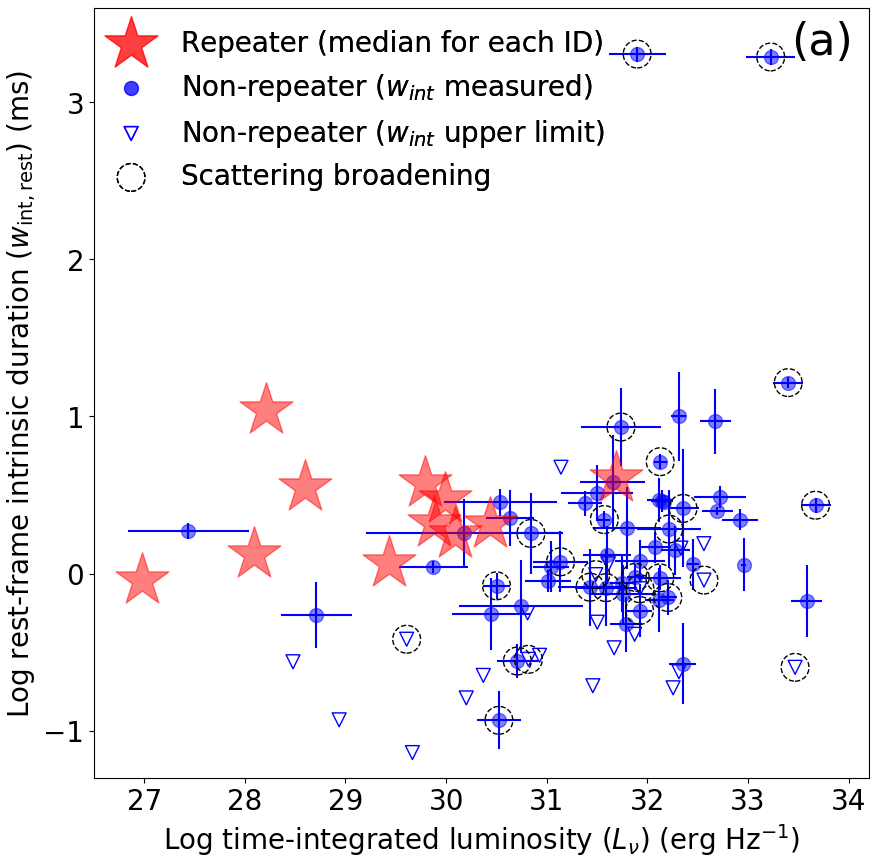}
	\includegraphics[width=\columnwidth]{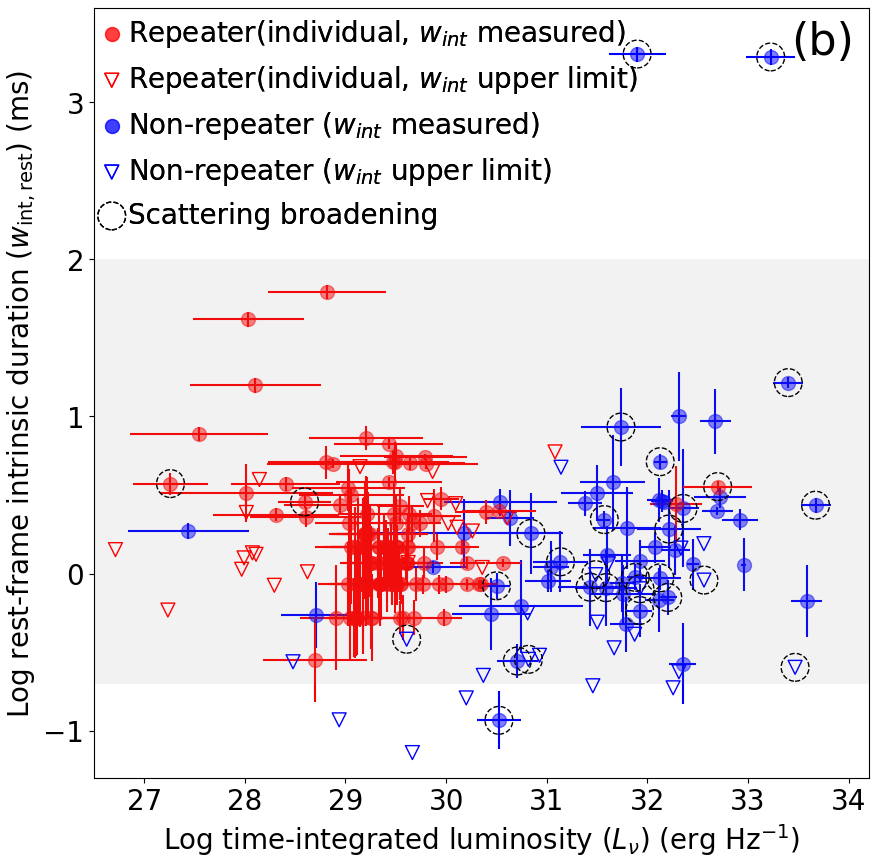}
    \caption{(Left) Rest-frame intrinsic duration as a function of time-integrated luminosity at rest-frame 1.83 GHz of FRBs.
    Red and blue colours are repeating and non-repeating FRBs, respectively.
    Each repeating FRB represents median values of the repeats.
    Upper limits on the rest-frame intrinsic duration are indicated by open triangles.
    FRBs with scattering broadening features are highlighted by dashed circles.
    Errors of $L_{\nu}$ and $w_{\rm int,rest}$ are calculated by Monte Carlo simulations.
    Random 10,000 errors are independently assigned to DM$_{\rm obs}$, $w_{\rm obs}$, DM$_{\rm IGM}(z)$, and $E_{\nu_{\rm obs}}$, which follow Gaussian probability distribution functions with the observational uncertainties.
    The line-of-sight fluctuation of DM$_{\rm IGM}(z)$ \citep{Zhu2018} is included in the error of DM$_{\rm IGM}(z)$.
    (Right) Same as left, except that the repeats of each repeating FRB are individually demonstrated.
    Grey shaded region is magnified in Fig. \ref{fig5}.
    }
    \label{fig4}
\end{figure*}

Fig. \ref{fig4} indicates rest-frame intrinsic duration, $w_{\rm int, rest}$, as a function of time-integrated luminosity, $L_{\nu}$, of FRBs.
Repeating and non-repeating FRBs are shown by red and blue colours, respectively.
In the left panel of Fig. \ref{fig4}, median values of repeating pulses of each FRB are shown in this parameter space (red stars), while individual repeats are shown in the right panel (red dots and triangles).
We found that repeating FRBs occupy relatively fainter and longer duration compared with non-repeating ones in the $L_{\nu}$-$w_{\rm int, rest}$ space (see also Fig. \ref{fig1}).
%KS tests for histograms of $L_{\nu}$ and $w_{\rm int, rest}$ are XXX, indicating statistically different distributions of repeating and non-repeating FRBs.

Fig. \ref{fig4} also indicates no clear correlation between $L_{\nu}$ and $w_{\rm int, rest}$ for repeating FRBs.
In contrast, we confirmed a positive correlation between $L_{\nu}$ and $w_{\rm int, rest}$ of non-repeating FRBs \citep{Hashimoto2019} except for some outliers with very long duration and faint luminosity.
These results suggest different origins of repeating and non-repeating FRBs (see Section \ref{discussion} for details).

\begin{figure*}
	\includegraphics[width=6.5in]{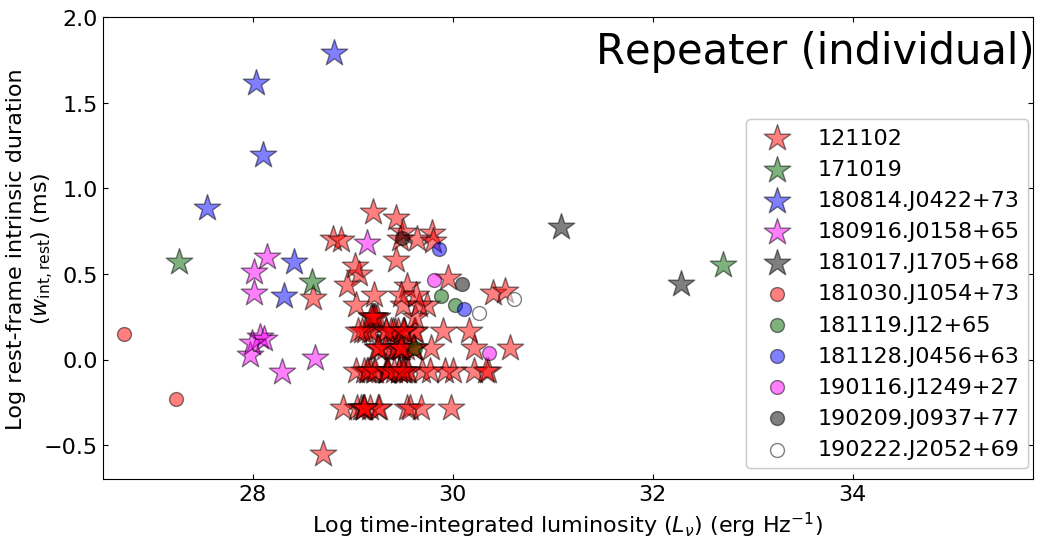}
    \caption{Magnified figure of the grey shaded region in Fig. \ref{fig4}(b).
    Only repeating FRBs are shown.
    Repeats from the same progenitor, i.e., the same FRB ID, are demonstrated by the same symbols listed in the legend.}
    \label{fig5}
\end{figure*}

\begin{figure*}
	\includegraphics[width=6.5in]{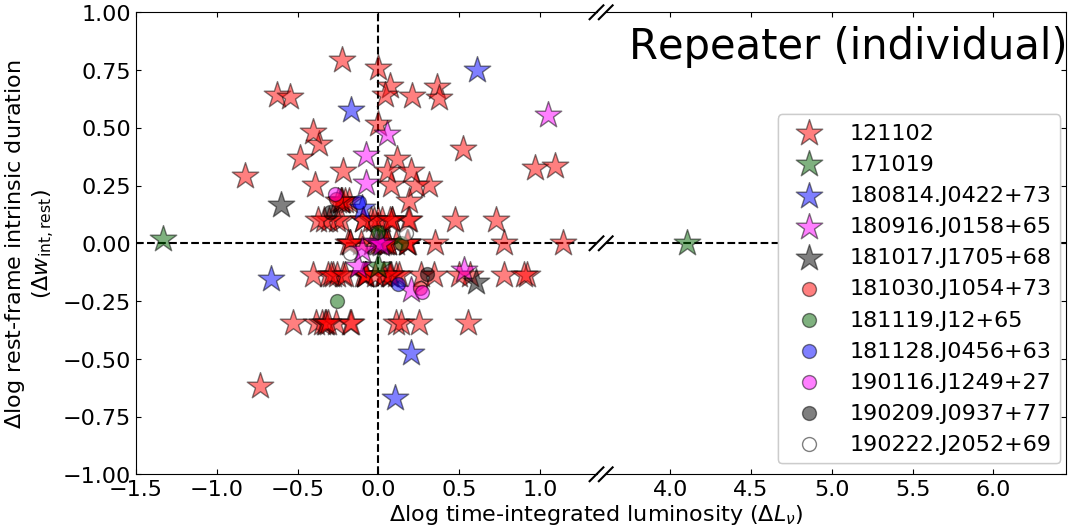}
    \caption{Same as Fig. \ref{fig5} except for off-set axes from median coordinates of each repeating FRB in the $L_{\nu}$-$w_{\rm int, rest}$ parameter space.
    }
    \label{fig6}
\end{figure*}

Distribution of repeating FRBs in a grey shaded region in Fig. \ref{fig4}b is magnified in Fig. \ref{fig5}.
Different markers correspond to different repeating FRBs in Fig \ref{fig5}.
During repeats of FRBs, they move around in this parameter space. 
However, we do not find any clear trend in terms of $L_{\nu}$ and $w_{\rm int, rest}$. 
This point is also confirmed in Fig. \ref{fig6} that is demonstrated in offset distances from median coordinates of each repeating FRB in the $L_{\nu}$-$w_{\rm int, rest}$ plane.

\subsection{Cadences of repeating FRBs as a function of luminosity}
\begin{figure*}
	\includegraphics[width=6.5in]{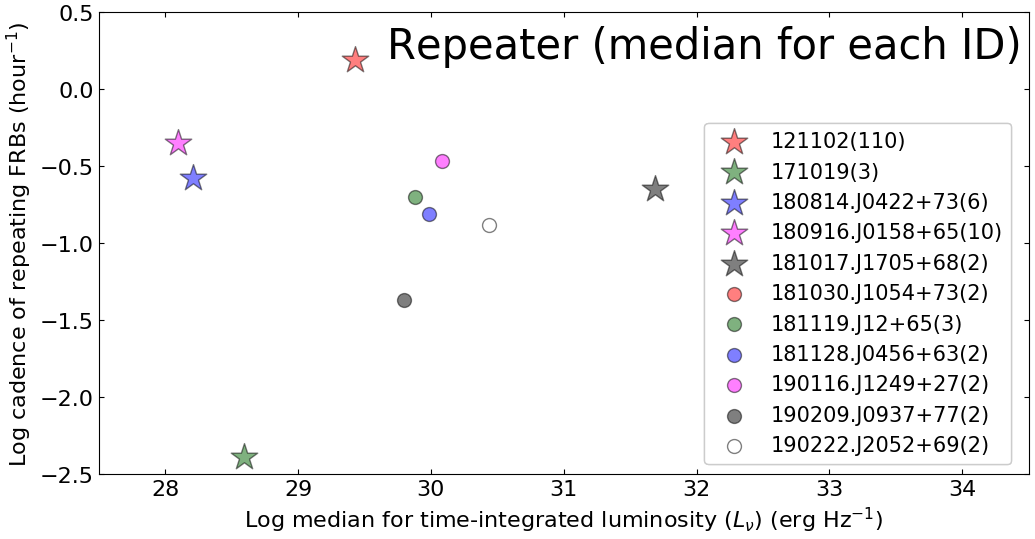}
    \caption{Averaged cadence as a function of median of time-integrated luminosity at rest-frame 1.83 GHz of repeating FRBs.
    Markers are the same as Fig \ref{fig6}. 
    Numbers in parentheses are the number of repeats for each repeating FRB.
    }
    \label{fig7}
\end{figure*}

We also investigated rest-frame cadences of repeating FRBs as a function of time-integrated luminosity in Fig. \ref{fig7}.
In Fig. \ref{fig7}, averaged repeating cadence of each repeating FRB is compared with median of $L_{\nu}$.
We found no clear correlation between these two parameters.

\subsection{Luminosity function of FRBs}
\begin{figure*}
	\includegraphics[width=5.2in]{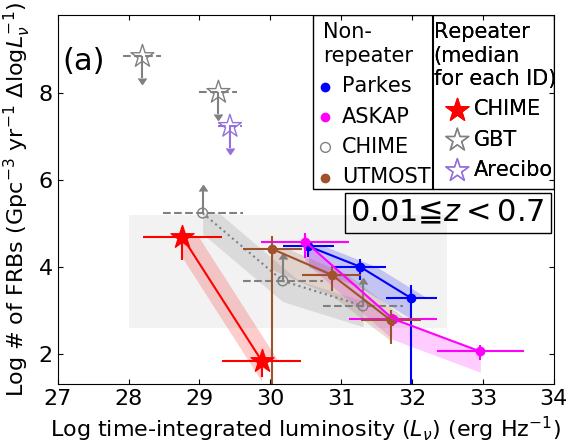}
	\includegraphics[width=5.2in]{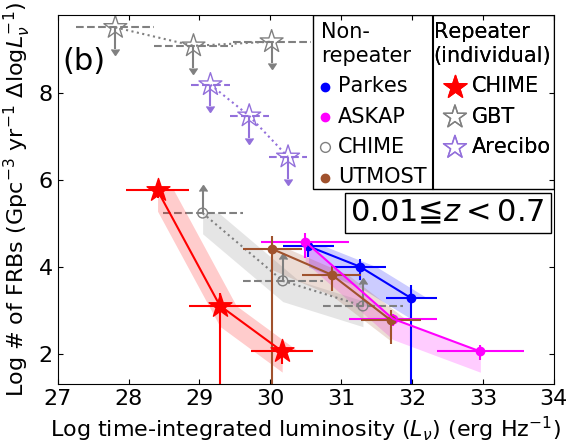}
    \caption{(Top) Luminosity function ($\log\Phi$) of FRBs.
    Different colours indicate different telescopes.
    The luminosity functions of repeating and non-repeating are shown by stars and dots, respectively.
    Repeating FRBs are counted such that the identical FRB ID is the single source.
    CHIME non-repeating FRBs place lower limits (grey open circles) because of the upper limit of the survey volume (Table \ref{tab3}). 
    GBT and Arecibo FRBs place upper limits (grey and purple stars, respectively) because of targeted observations of repeating FRBs. 
    Horizontal error bars correspond to luminosity bin sizes.
    Vertical error bars include the Poisson uncertainties.
    Upper shaded regions around the luminosity functions for Parkes, CHIME, and UTMOST correspond to systematic uncertainties in the time-integrated luminosity with respect to the unknown positions of FRBs within the fields of view. 
    Lower shaded regions correspond to systematic uncertainties of the effective survey areas arising from the uncertainty in the slope of source counts, $\alpha_{\rm SC}$.
    Grey square region is magnified in Fig. \ref{fig9}.
    (Bottom) Same as top except for the different method to count the number of repeating FRBs.
    The repeats of bursts are individually counted for each repeating FRB.
    }
    \label{fig8}
\end{figure*}

We show luminosity functions of repeating and non-repeating FRBs at $0.01 \leqq z<0.7$ in Fig. \ref{fig8}.
The luminosity functions are independently calculated for repeating (stars) and non-repeating (dots) FRBs detected by different telescopes because of different survey parameters as shown in Table \ref{tab2} and \ref{tab3}. 
The uncertainty in the time-integrated luminosity with respect to the unknown positions of FRBs is shown by upper shaded regions around the luminosity functions for Parkes, CHIME, and UTMOST.
Lower shaded regions correspond to systematic uncertainties of the effective survey areas arising from the uncertainty in the slope of source counts, $\alpha_{\rm SC}$ (see Section \ref{calc_LF} for details).

Since the FRB luminosity function is dependent on the number density of progenitors and the event rate, there are two possible ways to count the number of repeating FRBs. 
One is by counting number of progenitors, i.e., one FRB ID corresponds to one FRB.
Another is by counting number of repeats individually. 
The former is a fair way to count the number density of the progenitors unless repeating FRBs contaminate non-repeating sample significantly.
The latter could be a fairer way to count the event rate if the non-repeating FRBs are significantly contaminated by the repeating ones.
%This might be fair if only a part of repeats are detected by current telescopes and categorised into non-repeating FRBs. (XXX which is fair to what? --> done)

In the left panel of Fig. \ref{fig8}, the number of repeating FRBs are counted such that the identical FRB ID is the single source, while in the right panel, repeats are counted individually for each FRB.
GBT and Arecibo observations for repeating FRBs place upper limits on the luminosity functions (open grey and purple stars with arrows) since the observations were targeted on already known locations of repeating FRBs \citep{Spitler2014,Spitler2016,Scholz2016,Zhang2018,Kumar2019}.
CHIME observations for non-repeating FRBs place lower limits (open grey dots with arrows) since only the upper limit on the survey volume is provided due to the pre-commissioning operation \citep{CHIMEFRB2019}.

In Fig. \ref{fig8} left, in spite of different sensitivities and survey parameters, different telescopes provide roughly similar luminosity functions for non-repeating FRBs. 
The lower limits from CHIME non-repeating FRBs are consistent with other non-repeating luminosity functions.

The CHIME luminosity function of repeating FRBs in the left panel of Fig. \ref{fig8} (red stars) clearly deviates from luminosity functions of non-repeating FRBs.
This deviation is still obvious when the repeats are counted individually (right panel in Fig \ref{fig8}). 
%Therefore the luminosity functions imply different populations of repeating and non-repeating FRBs (see Section \ref{discussion} for details). 
The difference in the luminosity functions imply that repeating and non-repeating FRBs are indeed different populations.
In Fig. \ref{fig8}, we also found that the volumetric occurrence rates strongly depend on luminosity of FRBs regardless of the repetition.
Bright FRBs are extremely rare compared to faint ones.

\section{Discussion}
\label{discussion}
\subsection{Luminosity function of FRBs in a previous study}
\citet{Luo2018} calculated normalised luminosity functions of FRBs with assumptions of different types of host galaxies and DM$_{\rm MW}$ models by \citet{Cordes2002} and \citet{Yao2017}.
Their sample includes one repeating FRB 121102 and 32 non-repeating FRBs at $0.13 <z<2.0$.
They used isotropic luminosity integrated over the frequency in units of erg s$^{-1}$.
In this work, a total of 7 repeating FRBs and 46 non-repeating FRBs are used for the calculations of the luminosity functions.
Our sample is limited to FRBs at $0.01 \leqq z<0.7$.
The upper bound could reduce the possible effect of the redshift evolution (e.g., Hashimoto et al. 2020b in prep.).
The lower bound could reduce the large uncertainty on distances to nearby FRBs.
We use the time-integrated luminosity in units of erg Hz$^{-1}$ without integration over the frequency.
Due to these differences, the direct comparison between luminosity functions by \citet{Luo2018} and this work is not straightforward.
\citet{Luo2018} reported a power-law slope of the normalised luminosity functions ranging from $\alpha_{\rm LF}= -1.8$ to $-1.2$.
We fitted the luminosity functions with power-law functions weighted by the inverted Poisson errors. 
The best-fit power-law slopes are $\alpha_{\rm LF}=-2.5, -2.1$, and $-1.0$ for repeating FRBs (CHIME median for each ID), repeating FRBs (CHIME individual count), and non-repeating FRBs including Parkes, ASKAP, and UTMOST, respectively.
Although \citet{Luo2018} reported a cutoff of the luminosity function at the bright end, we do not find any clear cutoff up to $\sim10^{34}$ erg Hz$^{-1}$.
This is probably because we do not assume any functional shape to calculate the luminosity functions.
However, they assumed a Schechter luminosity function that includes a cutoff.
We do not exclude the possibility of a cutoff beyond $10^{34}$ erg Hz$^{-1}$, where no FRB has been found in our sample.
\citet{Shannon2018} suggested either a decreasing number of FRBs towards the bright end or a cutoff above $10^{34}$ erg Hz$^{-1}$ on the basis of ASKAP fly's-eye survey data. 
The luminosity functions of non-repeating FRBs in this work (Fig. \ref{fig8}) decline towards the bright end of $\sim 10^{34}$ erg Hz$^{-1}$, confirming the point presented by \citet{Shannon2018} with better statistics including FRBs detected with other telescopes.

\subsection{Astrophysical implications on FRB populations}
Observationally, FRBs are divided into two categories: repeating and non-repeating.
However, if only a single burst is detected among repeats due to low telescope sensitivity or long period between repeats, it may be recognised as a non-repeating FRB.
Therefore non-repeating FRB could be significantly contaminated by repeating FRBs.
In this sense these two categories do not necessarily indicate two different origins.

\citet{Ravi2019repeat} estimated lower limits on a volumetric occurrence rate of non-repeating FRBs detected by CHIME during the pre-commissioning phase.
The lower limits actually depend on the assumed dispersion measures of host galaxies.
In many cases, the lower limits exceed volumetric occurrence rates of possible progenitors of non-repeating FRBs, e.g., neutron-star merger and white-dwarf merger.
This suggests that at least some fractions of non-repeating FRBs originate from sources that emit repeating bursts over their lifetimes.

We found that repeating and non-repeating FRBs occupy different parameter spaces in the $L_{\nu}$-$w_{\rm int, rest}$ plane in Fig. \ref{fig4}.
Although several repeating FRBs overlap with non-repeating FRBs and vice versa, majorities of FRBs are clearly separated in Fig. \ref{fig4}. 
Repeating FRBs show relatively longer $w_{\rm int, rest}$ on average and much fainter $L_{\nu}$ compared with those of non-repeating FRBs (Fig. \ref{fig4} left).
These differences are also demonstrated in Figs. \ref{fig1}(j) and (k).
The cumulative histograms of repeating FRBs (red solid and dashed lines in Fig. \ref{fig1}) indicate longer $w_{\rm int, rest}$ and fainter $L_{\nu}$ than those of non-repeating FRBs (blue solid lines in Fig. \ref{fig1}).
We note that the difference in $w_{\rm int, rest}$ is marginal, when individual repeats of repeating FRBs are compared with that of non-repeating FRBs.
A difference in observed duration between repeating and non-repeating FRBs was reported by \citet{CHIME8repeat2019}.
We confirmed it with a more physically motivated parameter, $w_{\rm int, rest}$.
%The time-integrated luminosity difference is more significant than that of duration.
The difference between the repeating and non-repeating FRBs is more obvious in terms of time-integrated luminosity than the rest-frame duration.
%Fig. \ref{fig4} also indicates that there is no clear correlation for repeating FRBs in contrast to a positive correlation \citep{Hashimoto2019} for repeating FRBs except for a few outliers.

Upper limits on $w_{\rm int,rest}$ are shown by triangles in Fig. \ref{fig4}. 
Since these FRBs are not temporally resolved, they might include scattered pulses if observed with a much higher time resolution. 
These FRBs could have much shorter $w_{\rm int,rest}$ than the upper limit after removing the scattering component.
Therefore, several temporally unresolved repeating and non-repeating FRBs (red and blue triangles in Fig. \ref{fig4}b, respectively) might actually overlap.
%However, such overlapping FRBs can not be a majority, since the repeating and non-repeating FRBs have different time-integrated luminosities. 
However, the overall distributions of repeating and non-repeating FRBs in Fig. \ref{fig4}b are different, since they have different time-integrated luminosities.
In terms of two distinct FRB populations in the $L_{\rm nu}$-$w_{\rm int,rest}$ space, the potential contributions of scattering broadening to the temporally unresolved bursts do not significantly affect our argument. 

In Fig. \ref{fig4}, there is no clear correlation between the rest-frame duration and luminosity for repeating FRBs. 
This is in contrast to the positive correlation found for non-repeating FRBs \citep{Hashimoto2019}.
The time-integrated luminosity, i.e., a total isotropic energy released by the FRB, and duration are physically different quantities. 
Therefore, it is not necessary for these two quantities to be correlated even though the former is integrated over the duration. 
For instance, in the case of GRBs,
similar parameters, $E_{\rm iso}$ and $T_{90}$, have been investigated in literature \citep[e.g., ][]{Pelangeon2008}. 
Here, $E_{\rm iso}$ is the time-integrated luminosity of gamma-rays and $T_{90}$ is the duration which includes 90\% of the total gamma-ray fluence. 
These two physical parameters of GRBs do not show any clear correlation \citep{Pelangeon2008}. 
Similarly, the correlation seen in the non-repeating FRB relation is not an artificial consequence from having \lq time\rq\ in the two quantities. 
The detection limit depending on the observed duration, i.e., $E_{\rm lim} \propto w_{\rm obs}^{1/2}$, does not mimic the observed correlation (see Appendix C).
Whether or not these two parameters correlate depends on the FRB models (see Section \ref{FRBmodel} for details).
Different data distributions and non-correlation/correlation in the $L_{\nu}$-$w_{\rm int,rest}$ space 
%These observational results
suggest different physical origins of repeating and non-repeating FRBs.

In Fig. \ref{fig5} there are, at least, two repeating FRBs which change the $L_{\nu}$ dramatically from luminous non-repeating regime to faint repeating one, i.e., FRB 171019 and 181017.J1705+68 \citep{Kumar2019,CHIME8repeat2019}.
\cite{Kumar2019} reported repetitions of FRB 171019 which are 590 times fainter than the one discovered by ASKAP (brightest burst in Fig. \ref{fig5}) at different observed frequencies.
In this work, the time-integrated luminosity is compared at the same frequency by assuming a spectral index measured for each repeat \citep{Kumar2019}.
Therefore the difference in $L_{\nu}$ between the bright and faint repeats of FRB 171019 are larger than the factor of 590.
Repeats of these two FRBs originate from the same progenitors in spite of the large $L_{\nu}$ differences.
Therefore our results do not rule out a possibility that some of non-repeating FRBs are actually repeating but only the luminous repeats were detected because of sensitivity limits of telescopes.

Fig. \ref{fig8} demonstrates that luminosity functions of repeating FRBs are clearly different from that of non-repeating FRBs.
Even in the case of individual counting of repeats, the luminosity function of repeating FRBs is $\sim$ 2 order of magnitude lower than that of non-repeating FRBs.
The luminosity function would support the hypothesis of the different populations of repeating and non-repeating FRBs.
We note that $\sim$ 50\% contamination of repeating FRBs in the non-repeating sample could mitigate the difference, since the 50\% contamination decreases the number of non-repeating FRBs by 50\% and increases the number of repeating FRBs up to the same level.

\subsection{Implications on FRB models}
\label{FRBmodel}
A number of physical models of repeating and non-repeating FRBs have been proposed \citep[e.g., ][]{Platts2019}. 
Models of repeating FRBs include a neutron star-white dwarf (NS-WD) accretion \citep{Gu2016}, binary neutron-star mergers \citep[e.g.,][]{Yamasaki2018}, active galactic nuclei (AGN)-compact object interaction \citep[e.g., ][]{Gupta2018}, AGN jet \citep[e.g., ][]{Katz2017}, NS-asteroid belt interaction \citep{Dai2016}, magnetar \citep[e.g., ][]{Beloborodov2017,Margalit2019,Wadiasingh2019,Metzger2019}, pulsar lightning \citep{Katz2017light}, starquake \citep{Wang2018,Suvorov2019}, wandering pulsar beam \citep{Katz2017beam}, and giant pulse of a young pulsar \citep[e.g.,][]{Keane2012,Cordes2016,Connor2016}.

Models of non-repeating FRBs include a collapse of a neutron star \citep[e.g., ][]{Fuller2015,Falcke2014,Shand2016}, NS-asteroid collision \citep{Geng2015}, pulsar-black hole (BH) interaction \citep{Bhattacharyya2017}, merger of compact objects \citep[e.g., ][]{Zhang2016,Liu2016,Mingarelli2015,Totani2013,Liu2018,Li2018,Kashiyama2013,Yamasaki2018}, NS-supernova (SN) interaction \citep{Egorov2009}, AGN jet-cloud interaction \citep[e.g., ][]{Romero2016} and SN remnant powered by a flare from a magnetar \citep[e.g., ][]{Popov2010,Lyubarsky2014,Murase2016}.

There is no conclusive consensus on the physical origins of FRBs so far.
One of the central foci of FRB studies is an observational constraint on the models.
Detailed comparisons between observational results and individual physical models are beyond the scope of this work.
Here we briefly sketch out rough constraints on the physical origins of repeating and non-repeating FRBs implied from observational results shown in Section \ref{results}.

The $L_{\nu}$-$w_{\rm int, rest}$ relation is one way to constrain the models. 
The observed positive correlation between $L_{\nu}$ and $w_{\rm int, rest}$ of non-repeating FRBs could favour scenarios which predict the positive correlation with slopes similar to the observed value \citep{Hashimoto2019}.  
The favoured scenarios for non-repeating FRBs are, at least, (i) AGN-jet cloud interaction \citep[e.g., ][]{Romero2016}, (ii) NS-asteroid collision \citep[e.g., ][]{Geng2015}, and (iii) SN remnant powered by a magnetar \citep[e.g., ][]{Lyubarsky2014}, since these scenarios predict a positive correlation between $L_{\nu}$ and $w_{\rm int, rest}$ \citep[see ][for details]{Hashimoto2019}.
%In this sense no clear correlation of repeating FRBs shown in Fig. \ref{fig4} to \ref{fig6} would also be a hint to understand their origin.
As there is no clear correlation for repeating FRBs shown in Fig. \ref{fig4} to \ref{fig6}, a possible model to explain this non-correlation is e.g. pulsar lightning \citep{Katz2017light}.
%Among the physical models of repeating FRBs, models which do not explicitly predict the correlation between $L_{\nu}$ and $w_{\rm int, rest}$ would be preferential, e.g., pulsar lightning \citep{Katz2017light}. 

Repeating period is another hint to constrain the repeating-FRB models.
If repeating FRBs are triggered by accretion of materials \citep[e.g., ][]{Gu2016,Katz2017}, higher energy or brighter luminosity likely requires longer accretion time to accumulate more materials, i.e., longer period.
There is no clear correlation showed in Fig. \ref{fig7}, though the numbers of FRBs and repeats are very small.
%This might imply origins except for accretion phenomena.
%We need to wait for future data to conclude this point.
This may rule out accretion as a possible origin mechanism of repeating FRBs. However, more future data is needed to ascertain this point.

A volumetric occurrence rate of FRBs could also provide important indication on FRB progenitors.
\citet{Ravi2019repeat} calculated the local volumetric occurrence rate of non-repeating FRBs based on a distance to the second or third closest FRBs detected by CHIME.
The rate is a lower limit because there might be other faint FRBs under the detection limit within the distance.
\citet{Ravi2019repeat} demonstrated that the lower limit is higher than the number density of possible progenitors of FRBs and argued that most cases of non-repeating FRBs should repeat during the lifetime of the progenitors.
However, only nearby two or three FRBs are used in the analysis in spite of the detection of other FRBs.
The distance to the nearby FRBs is relatively more uncertain compared to that of distant FRBs because DM$_{\rm IGM}$ is used as a distance indicator.
The DM contamination from the Milky Way and the host galaxies become relatively larger for closer FRBs, which makes the distance to the nearby FRBs and the volumetric density more uncertain.
The volumetric occurrence rate in \citet{Ravi2019repeat} is also very sensitive to the peculiarities of the few lowest-DM events.
In addition, the number density of FRB likely depends on the luminosity as other sources in the Universe do.

\begin{figure*}
	\includegraphics[width=6.5in]{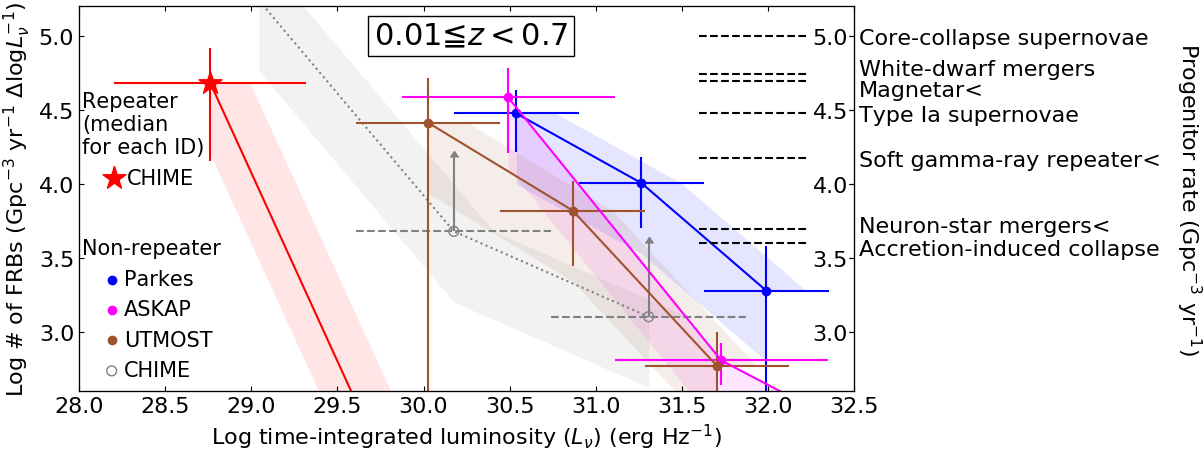}
    \caption{Magnified figure of the grey square region in Fig. \ref{fig8} (a), comparing with possible progenitors' volumetric occurrence rates \citep{Ofek2007,Keane2008,Li2011,Badenes2012,Taylor2014,Abbott2017,Ruiter2019,Ravi2019repeat}.
    The rates of magnetar, soft gamma-ray repeater, and neutron-star mergers correspond to the upper limits \citep{Ofek2007,Keane2008,Abbott2017}.
    Repeating FRBs are counted such that the identical FRB ID is the single source.
    Upper shaded regions around the luminosity functions for Parkes, CHIME, and UTMOST correspond to systematic uncertainties with respect to the unknown positions of FRBs within the fields of view. 
    Lower shaded regions correspond to systematic uncertainties of the effective survey areas arising from the uncertainty in the slope of source counts, $\alpha_{\rm SC}$.
    }
    \label{fig9}
\end{figure*}

In this work we used the $V_{\rm max}$ method to calculate volumetric occurrence rate as a function of luminosity, i.e., luminosity function of FRBs. 
The luminosity function contains the statistics of all FRBs at $0.01\leqq z<0.7$ detected with each telescope, taking the detection limit and luminosity into account.
In Fig. \ref{fig8}, the luminosity functions of CHIME-detected non-repeating FRBs (grey dots with arrows) are still lower limits because only the upper limit on the survey area of CHIME pre-commissioning observations is provided in the literature \citep{CHIMEFRB2019}.
The grey square region in the left panel of Fig. \ref{fig8} is magnified in Fig. \ref{fig9} together with volumetric occurrence rates of possible progenitors \citep{Ofek2007,Keane2008,Li2011,Badenes2012,Taylor2014,Abbott2017,Ruiter2019,Ravi2019repeat}.
Note that the volumetric occurrence rate of each possible progenitor is integrated over luminosity.
The integration of the FRB luminosity function is almost determined by the faintest bin. 

%In Fig. \ref{fig9}, the faint end of the luminosity function of the repeating FRBs detected by CHIME (red star) is comparable to those of neutron-star mergers \citep{Abbott2017} and accretion-induced collapse of white dwarfs \citep{Moriya2016,Ruiter2019}.
In Fig. \ref{fig9}, the faint end of the luminosity function of the repeating FRBs detected by CHIME (red star) is comparable to those of white-dwarf mergers \citep{Badenes2012}, magnetars \citep{Keane2008}, type Ia supernovae \citep{Li2011}, and soft gamma-ray repeaters \citep{Ofek2007}, indicating that faint repeating FRBs may be related to these progenitors.
%neutron-star mergers \citep{Abbott2017} and accretion-induced collapse of white dwarfs \citep{Moriya2016,Ruiter2019}.
The bright end is lower than any of the possible candidates of the progenitors.
The bright population of repeating FRBs is very rare, suggesting that it occurs only in an extremely small portion of the progenitors.

%The faint ends of the luminosity functions of non-repeating FRBs (blue, magenta, and brown dots) are beyond neutron-star mergers and accretion-induced collapse of white dwarfs.
%Here there are two possibilities, i.e., either (i) faint non-repeating FRBs originate in neutron-star mergers or accretion-induced collapse and are actually repeating during the lifetime of the progenitor or (ii) faint non-repeating FRBs do not originate in these possible progenitors.
%The case (i) is consistent with the discussion presented in \citet{Ravi2019repeat}.
%There might be contamination of repeating FRBs in non-repeating ones.
%The case (ii) could indicate the progenitors of soft gamma-ray repeater \citep[e.g., ][]{Ofek2007} for faint non-repeating FRBs in terms of the volumetric occurrence rate.
%The bright end of the luminosity function of the non-repeating FRBs is lower than those of the possible progenitors.
%The bright non-repeating FRBs also might originate in a very small fraction of the possible progenitors.
%Otherwise, bright FRBs might originate in unknown progenitors.

The faint ends of the luminosity functions of non-repeating FRBs detected with Parkes, ASKAP, and UTMOST (faint ends of blue, magenta, and brown dots in Fig. \ref{fig9}) are comparable to that of repeating FRBs (red star) and thus have similar rates to white-dwarf mergers \citep{Badenes2012}, magnetars \citep{Keane2008}, type Ia supernovae \citep{Li2011}, and soft gamma-ray repeaters \citep{Ofek2007}.
The faint ends are higher than those of neutron-star mergers \citep{Abbott2017} and accretion-induced collapse of white dwarfs \citep{Moriya2016,Ruiter2019}.
There are two possibilities, i.e., either (i) faint non-repeating FRBs originate in neutron-star mergers or accretion-induced collapse and are actually repeating during the lifetime of the progenitor or (ii) faint non-repeating FRBs are 'truly non-repeating' without significant contamination from repeaters and do not originate in either neutron-star mergers or accretion-induced collapse.
The case (i) is consistent with the discussion presented in \citet{Ravi2019repeat}.
There might be contamination of repeating FRBs in the non-repeating ones.
The case (ii) could indicate that the progenitors of faint non-repeating FRBs are any of soft gamma-ray repeaters, type Ia supernovae, magnetars, and white-dwarf mergers in terms of the volumetric occurrence rate.
The bright end of the luminosity function of the non-repeating FRBs is lower than any of the possible candidates of the progenitors.
The bright non-repeating FRBs might also originate in a very small fraction of the possible progenitors.
Otherwise, bright FRBs might originate in unknown progenitors.

These statistical arguments are based on only the rate matching analysis.
More direct evidences would be obtained via follow-up observations of the possible progenitors. 
Although radio follow-up observations of the possible progenitors, e.g., GRBs \citep{Madison2019,Men2019} and supernova remnants \citep{Law2019} have not yet detected any FRB counterparts so far, direct detection of FRBs from them are necessary to give a more comprehensive answer to the FRB origins.

\section{Conclusions}
\label{conclusion}
We compiled a total of 11 repeating FRBs with 144 repeats and 77 non-repeating FRBs from the FRBCAT project.
From this sample we found that repeating and non-repeating FRBs are clearly distinguishable in the $L_{\nu}$-$w_{\rm int, rest}$ parameter space (Fig. \ref{fig4}), where $L_{\nu}$ and $w_{\rm int, rest}$ are the time-integrated luminosity and rest-frame intrinsic duration of FRBs, respectively.
%The repeating FRBs demonstrate relatively fainter $L_{\nu}$ and longer $w_{\rm int, rest}$ than those of the non-repeating FRBs.
The repeating FRBs have fainter $L_{\nu}$ and longer $w_{\rm int, rest}$ compared to the non-repeating population.
In contrast to non-repeating FRBs, repeating FRBs do not show any clear correlation between the $L_{\nu}$ and $w_{\rm int, rest}$.
We also found that the luminosity function of repeating FRBs is much lower than that of non-repeating FRBs (Fig. \ref{fig8}).
These results imply that repeating and non-repeating FRBs are essentially different populations.

During the repeats of FRBs, each repeat randomly moves in the $L_{\nu}$-$w_{\rm int, rest}$ plane without any clear trend (Figs. \ref{fig5} and \ref{fig6}).
Physical models which do not predict any correlation between the $L_{\nu}$ and $w_{\rm int, rest}$ could be favoured for repeating FRBs, e.g., a pulsar lightning scenario.
Accretion-material scenario might be disfavoured for repeating FRBs, since we do not find any clear correlation between the cadence and time-integrated luminosity (Fig. \ref{fig7}).

%The faint end of the luminosity function of the repeating FRBs is consistent with the volumetric occurrence rates of neutron-star mergers and accretion-induced collapse of white dwarfs, while non-repeating one exceeds them (Fig. \ref{fig9}).
The faint ends of the luminosity functions of repeating and non-repeating FRBs are higher than volumetric occurrence rates of neutron-star mergers and accretion-induced collapse of white dwarfs (Fig. \ref{fig9}).
They are consistent with the rates of soft gamma-ray repeaters, type Ia supernovae, magnetars, and white-dwarf mergers.
This indicates two possibilities: either (i) faint non-repeating FRBs originate in neutron-star mergers or accretion-induced collapse and are actually repeating during the lifetime of the progenitor or (ii) faint non-repeating FRBs are not related to these possible progenitors but originate in any of soft gamma-ray repeaters, type Ia supernovae, magnetars, and white-dwarf mergers.

The bright ends of luminosity functions of repeating and non-repeating FRBs are lower than any candidates of progenitors (Fig. \ref{fig9}).
This suggests that bright FRBs are extremely rare and are produced from a very small fraction of the progenitors regardless of the repetition.
Otherwise, bright FRBs might originate in unknown progenitors.

\section*{Acknowledgements}
We are very grateful to the anonymous referee for many insightful comments.
TH and AYLO are supported by the Centre for Informatics and Computation in Astronomy (CICA) at National Tsing Hua University (NTHU) through a grant from the Ministry of Education of the Republic of China (Taiwan).
TG acknowledges the support by the Ministry of Science and Technology of Taiwan through grant 108-2628-M-007-004-MY3.
AYLO's visit to NTHU was supported by the Ministry of Science and Technology of the ROC (Taiwan) grant 105-2119-M-007-028-MY3, hosted by Prof. Albert Kong.
This work made use of the CICA Cluster at the NTHU/CICA, supported by the Taiwan Ministry of Education and NTHU.
This research has made use of NASA's Astrophysics Data System.

%%%%%%%%%%%%%%%%%%%%%%%%%%%%%%%%%%%%%%%%%%%%%%%%%%



\bibliographystyle{mnras}
\bibliography{repeat_FRB_mnras} % if your bibtex file is called example.bib

%\begin{thebibliography}{99}
%\end{thebibliography}

%%%%%%%%%%%%%%%%%%%%%%%%%%%%%%%%%%%%%%%%%%%%%%%%%%

%%%%%%%%%%%%%%%%% APPENDICES %%%%%%%%%%%%%%%%%%%%%

\appendix
\section{DM$_{\rm IGM}$ and redshift distribution in each telescope}
\begin{figure*}
	\includegraphics[width=6.0in]{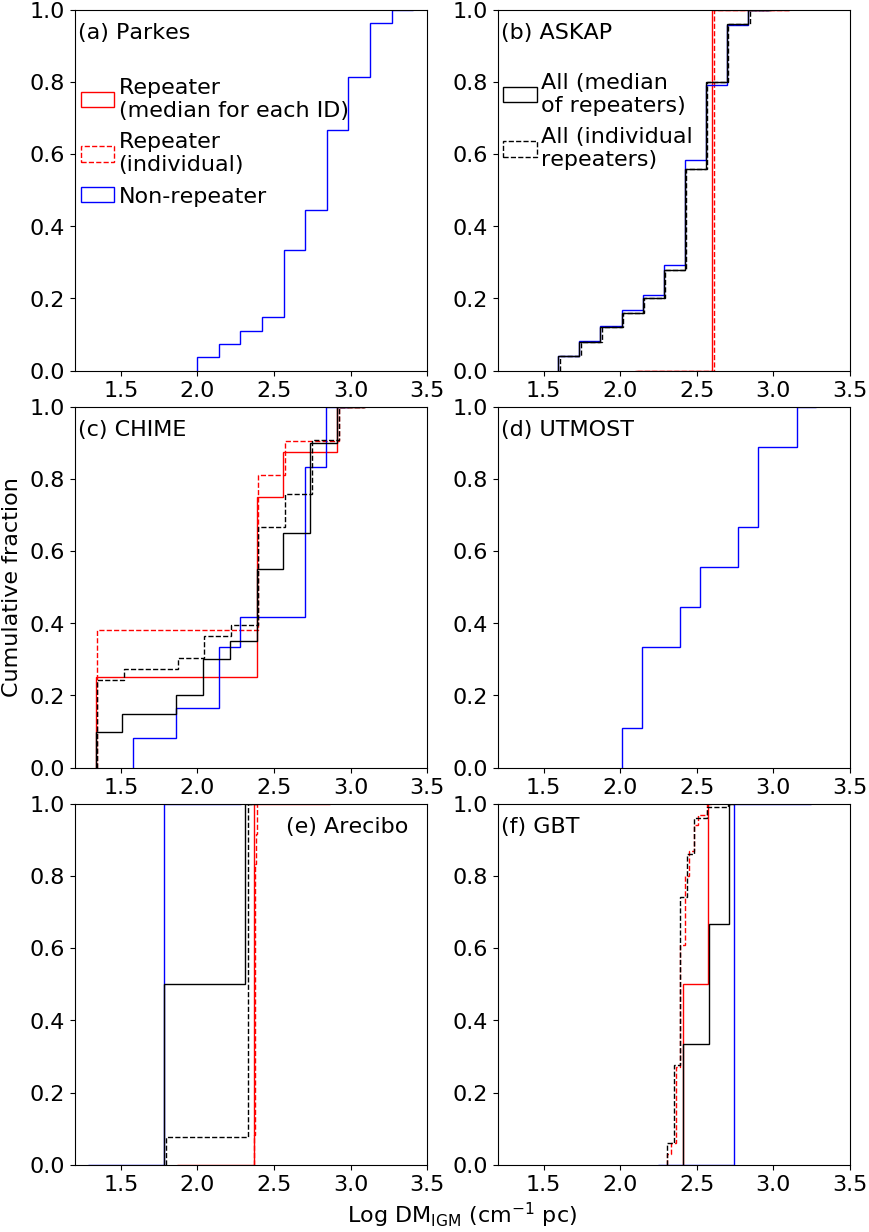}
    \caption{
    Same as Fig. \ref{fig1}e except for plotting samples of individual telescopes.
    }
    \label{figA1}
\end{figure*}

\begin{figure*}
	\includegraphics[width=6.0in]{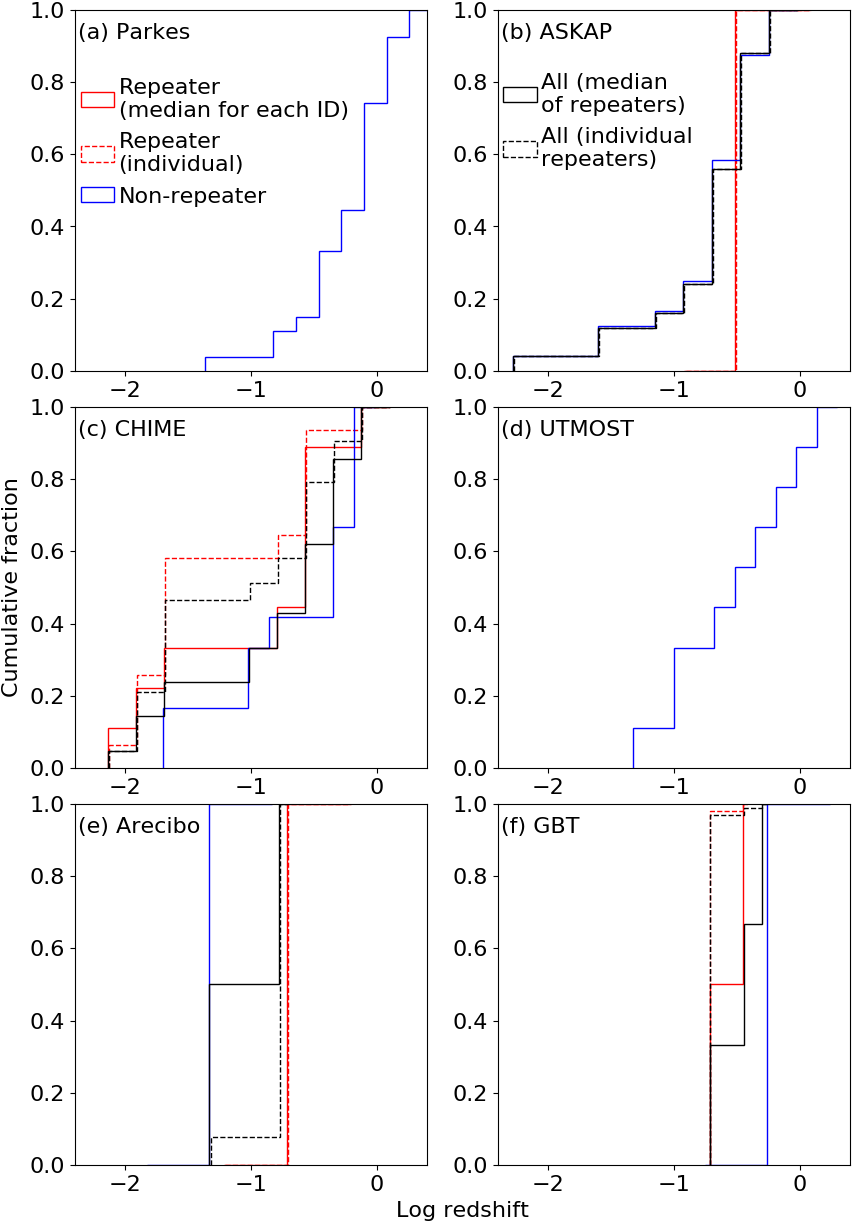}
    \caption{
    Same as Fig. \ref{fig1}f except for plotting samples of individual telescopes.
    }
    \label{figA2}
\end{figure*}

Fig. \ref{fig1} includes the contributions from a variety of telescopes with different survey sensitivities and DM distributions.
In Figs. \ref{figA1} and \ref{figA2}, we show the distributions of DM$_{\rm IGM}$ and redshift of each telescope in our sample, respectively.

\section{Detection limits depending on duration}
\begin{figure*}
	\includegraphics[width=6.0in]{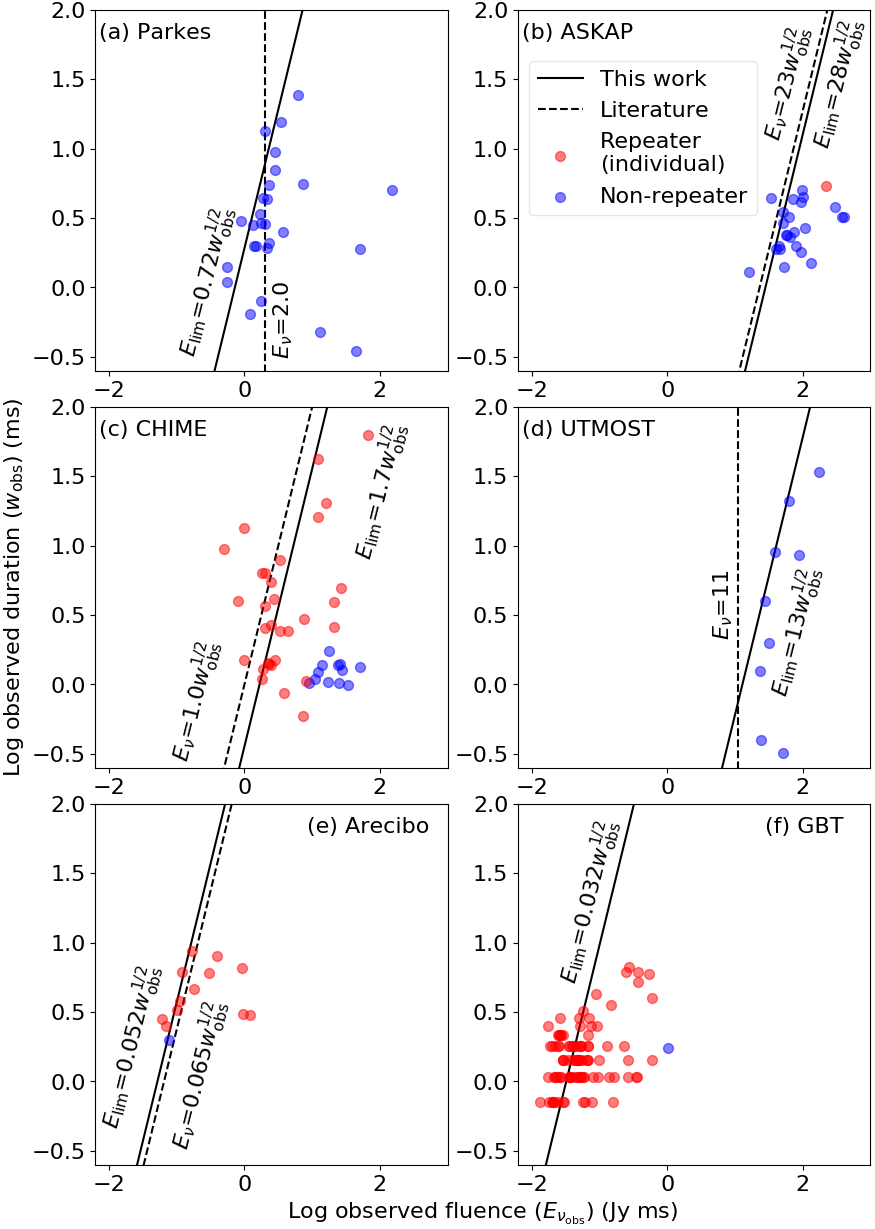}
    \caption{
    Observed duration as a function of fluence of our FRB sample.
    Each panel indicates data obtained by each telescope.
    Red and blue dots correspond to repeating and non-repeating FRBs, respectively. 
    Solid lines are detection limits that we empirically calculated (see also Fig. \ref{figB2}).
    Dashed lines are detection limits reported in literature \citep{Spitler2014,Keane2015,Caleb2016,Shannon2018,CHIMEFRB2019}.
    All of our sample is demonstrated before applying the redshift cuts and detection limits for the calculations of luminosity functions.
    }
    \label{figB1}
\end{figure*}

\begin{figure*}
	\includegraphics[width=6.0in]{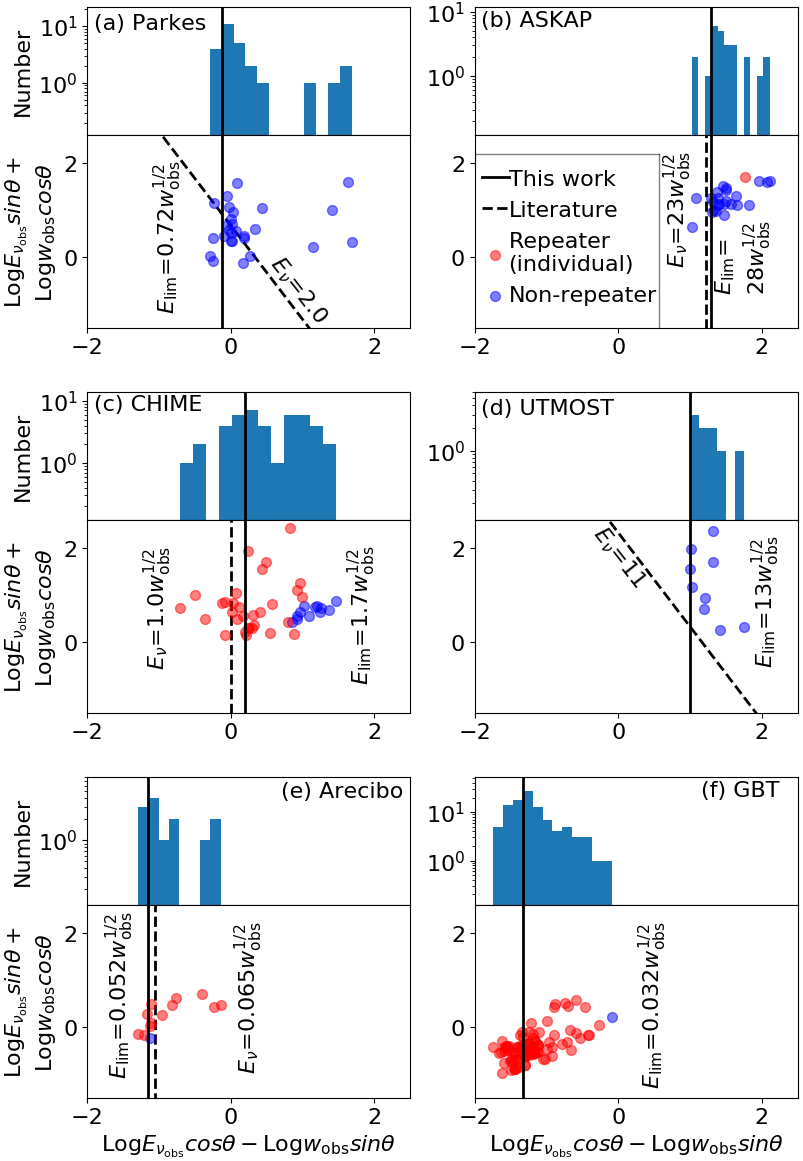}
    \caption{
    Same as Fig. \ref{figB1} except for rotation of the axes.
    The axes are rotated so that the duration dependency on the detection limit can be aligned along the vertical axis.
    Top figure in each panel shows a histogram of FRBs including repeating and non-repeating ones detected with each telescope.
    The rotation angle, $\theta$, is $\pi/2-\tan^{-1}(2.0)$ rad.
    The detection limits in our analysis are determined by the peaks of the histograms.
    }
    \label{figB2}
\end{figure*}

Here we empirically estimate a detection limit, $E_{\rm lim}$, for each telescope.
Fig. \ref{figB1} shows the observed duration as a function of observed fluence.
Each panel indicates FRBs detected with each telescope.
Dashed lines correspond to detection limits reported in previous works \citep{Spitler2014,Keane2015,Caleb2016,Shannon2018,CHIMEFRB2019}.
The reported detection limits of ASKAP, CHIME, and Arecibo include the duration dependency explicitly, while those of Parkes and UTMOST do not.
The different definitions of the detection limit could introduce additional systematics in different telescopes.
To remove such systematics, we adopt the same definition of the detection limit that involves the $w_{\rm obs}^{1/2}$ dependency.
For each telescope, we approximated $E_{\rm lim}$ as a peak of data distribution along the perpendicular direction to the $w_{\rm obs}^{1/2}$ dependency in Fig. \ref{figB1}.
The peak is utilised to reduce the uncertainty with respect to observational incompleteness.
In order to investigate the peak of data distribution, Fig. \ref{figB1} was rotated so that the $w_{\rm obs}^{1/2}$ dependency can be aligned along the vertical axis (Fig. \ref{figB2}).
The peaks of histograms in Fig. \ref{figB2} correspond to empirically determined $E_{\rm lim}$ as shown by the black solid lines in Figs. \ref{figB1} and \ref{figB2}.

\section{Does detection limit mimic the luminosity-duration relation?}
We performed Monte Carlo simulations to confirm whether the detection limits of radio telescopes, $E_{\rm lim} \propto w_{\rm obs}^1/2$, mimic the $L_{\nu}$-$w_{\rm int,rest}$ correlation found for non-repeating FRBs. 
In each simulation, we assumed 120 artificial data uniformly-distributed in the $L_{\nu}$-$w_{\rm int,rest}$ plane ranging from $\log(L_{\nu}) $= 30.0 to 34.0 (erg Hz$^{-1}$) and $\log(w_{\rm int, rest})=-1.0$ to 1.5 (ms). 
%This \lq intrinsic\rq\ data distribution is supposed to show no correlation 
This artificial data distribution has no \lq intrinsic\rq\ correlation, 
but the correlation coefficient slightly fluctuates due to the random process.
For the artificial redshift distribution, we used the best-fit Gaussian function to the redshift distribution of FRBs detected with Parkes, i.e., a Gaussian function centred at $z=0.86$ with $\sigma=0.6$. 
The redshift is randomly assigned to each artificial data following this Gaussian probability distribution function. 
We also randomly assigned galactic coordinates, $b$ and $l$, to each data because DM$_{\rm MW}$ and thus DM$_{\rm obs}$ affects $w_{\rm obs}$ through the dispersion smearing. 
%Among 120 artificial data, four telescopes including Parkes, ASKAP, CHIME and UTMOST are supposed to observe each 30 artificial data. 
Among 120 artificial data, each of the four telescopes -Parkes, ASKAP, CHIME, and UTMOST- will observed 30 artificial data points.
For each telescope, median values of $w_{\rm sample}$, $\nu_{\rm obs}$, and $\Delta\nu_{\rm obs}$ in our sample are utilised, where $w_{\rm sample}$, $\nu_{\rm obs}$, and $\Delta\nu_{\rm obs}$ are the sampling time, observed frequency, and intra-channel bandwidth, respectively. 
Based on these assumptions, we calculated the observed quantities, $E_{\nu_{\rm obs}}$ and $w_{\rm obs}$, %in reverse order of calculations 
following the methods described in Sections \ref{calcLnu} and \ref{calcw_int_rest}. 
We here define \lq detected FRBs\rq\ if $E_{\nu_{\rm obs}}$ is higher than $E_{\rm lim} (\propto w_{\rm obs}^{1/2}$). 
$E_{\rm lim}$ is empirically determined for each telescope (see Appendix B). 

\begin{figure}
	\includegraphics[width=\columnwidth]{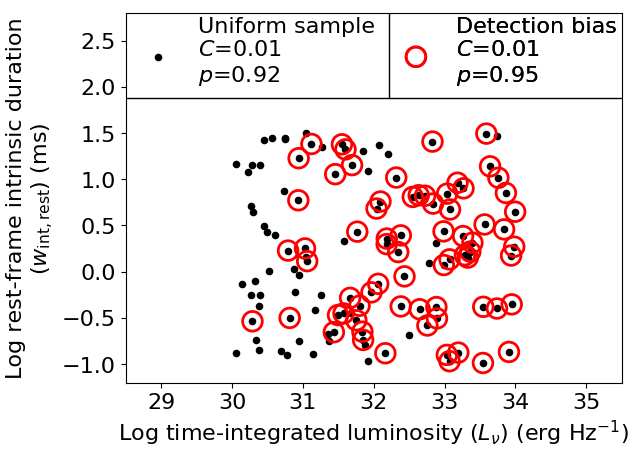}
    \caption{
    One example of the Monte Carlo simulations.
    Black dots are 120 simulated data following a uniform probability distribution in the $L_{\nu}$-$w_{\rm int, rest}$ plane.
    Four radio telescopes - Parkes, ASKAP, CHIME and UTMOST - are supposed to observe each 30 simulated data based on our detection limits (see also Figs. \ref{figB1} and \ref{figB2}).
    The detected sources are marked by red circles.
    We iterated this process 10,000 times to derive histograms of Pearson coefficients ($C$) and $p$-values ($p$) in Fig. \ref{figC2}, where $C$ and $p$ indicate the strength of correlation and statistical significance, respectively.
    This example indicates no significant correlation between $L_{\nu}$ and $w_{\rm int, rest}$ in the intrinsic uniform data distribution and detected FRBs.
    }
    \label{figC1}
\end{figure}

%One example of a simulation with 120 artificial data is shown in 
Fig. \ref{figC1} shows an example of a simulation with 120 artificial data points.
In this example, a Pearson coefficient ($C$) and $p$-value ($p$) are $C=0.01$ and $p=0.92$ for the intrinsic uniform data distribution, 
%and
whereas $C=0.01$ and $p=0.95$ for the detected FRBs.
$C$ and $p$ indicate the strength of correlation and statistical significance, respectively.
These values indicate no significant correlation between $L_{\nu}$ and $w_{\rm int, rest}$ in the intrinsic uniform data distribution and detected FRBs. 

\begin{figure}
	\includegraphics[width=\columnwidth]{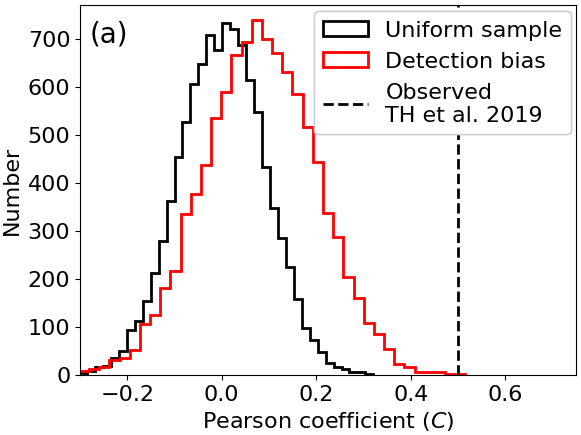}
	\includegraphics[width=\columnwidth]{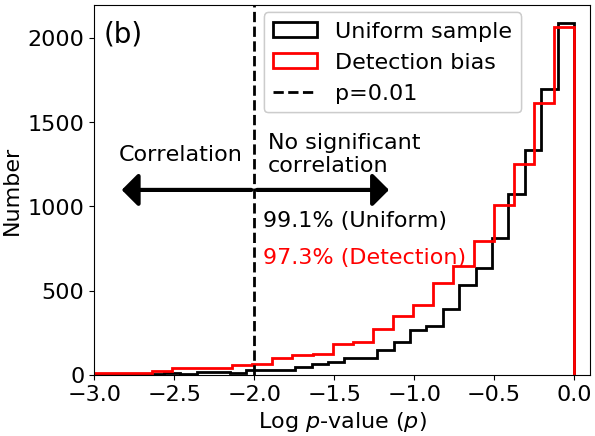}
    \caption{
    \label{figC2}Histograms of (Top) Pearson coefficients and (Bottom) $p$-values derived from the Monte Carlo simulation.
    The histograms of the original uniform sample are shown by black solid lines and those of detected sources are shown in red.
    Dashed vertical lines in the top and bottom panels correspond to a coefficient measured for the observed FRB sample in \citet{Hashimoto2019} and $p$=0.01, respectively. 
    }
\end{figure}

We iterated this 120-data simulation 10,000 times. 
The histograms of Pearson coefficients and $p$-values are shown in Fig. \ref{figC2}. 
In Fig. \ref{figC2}a, we found that the peak position of Pearson coefficients moved from $C=0.0$ to $\sim0.1$ after the detection limits are applied. 
This value of $C\sim0.1$ is much smaller than the observed one, i.e., $C\sim0.5$ \citep{Hashimoto2019}. 
%In Fig. \ref{figC2}b, a probability that detected FRBs indicate no significant correlation between $L_{\nu}$ and $w_{\rm int, rest}$ is 97.3\% among the 10,000 iterations. 
In Fig. \ref{figC2}b, 97.3\% of cases among 10,000 iterations indicate no significant correlation between $L_{\nu}$ and $w_{\rm int,rest}$. 
Therefore, we conclude that the duration-dependent detection limit does not mimic a statistically significant correlation. 

% Don't change these lines
\bsp	% typesetting comment
\label{lastpage}
\end{document}